\documentclass[sigconf, nonacm]{acmart}

\usepackage{libertine}
\usepackage{algorithm}

\usepackage{amsmath, amssymb, amsthm}
\usepackage[subrefformat=parens]{subcaption}
\usepackage{tikz}
\usepackage{xcolor}
\newtheorem{theorem}{Theorem}[section]
\newtheorem{lemma}{Lemma}[section]

\usepackage{multirow} 
\usepackage{amsthm}

\theoremstyle{remark}
\usepackage{subcaption, xspace}

\newcommand{\REM}[1]{}
\newcommand{\maxflow}{\textsf{MaxFlow}\xspace}

\usepackage{algpseudocode}
\algtext*{EndIf}
\newcommand\vldbdoi{XX.XX/XXX.XX}
\newcommand\vldbpages{XXX-XXX}
\newcommand\vldbvolume{19}
\newcommand\vldbissue{1}
\newcommand\vldbyear{2026}
\newcommand\vldbauthors{\authors}
\newcommand\vldbtitle{\shorttitle} 
\newcommand\vldbavailabilityurl{URL_TO_YOUR_ARTIFACTS}
\newcommand\vldbpagestyle{plain} 
\definecolor{tabred}{RGB}{214,39,40}
\definecolor{tabgreen}{RGB}{44,160,44}
\definecolor{tabblue}{RGB}{31,119,180}
\definecolor{tabpink}{RGB}{255,105,180} 

\begin{document}
\title{Efficient Dynamic \maxflow Computation on GPUs}


\author{Shruthi Kannappan}
\affiliation{%
  \institution{Indian Institute of Technology Madras}
  \city{Chennai}
  \country{India}
}
\email{shruthikannappan@gmail.com}
\author{Ashwina Kumar}
\affiliation{%
  \institution{Indian Institute of Technology Madras}
  \city{Chennai}
  \country{India}
}
\email{ashwinakumar825@gmail.com}
\author{Rupesh Nasre}
\affiliation{%
  \institution{Indian Institute of Technology Madras}
  \city{Chennai}
  \country{India}
}
\email{rupesh@cse.iitm.ac.in}

\begin{abstract}
\maxflow is a fundamental problem in graph theory and combinatorial optimisation, used to determine the maximum  flow from a source node to a sink node in a flow network. It finds applications in diverse domains, including computer networks, transportation, and image segmentation. The core idea is to maximise the total flow across the network without violating capacity constraints on edges and ensuring flow conservation at intermediate nodes. The rapid growth of unstructured and semi-structured data has motivated the development of parallel solutions to compute MaxFlow. However, due to the higher computational complexity, computing \maxflow for real-world graphs is time-consuming in practice. In addition, these graphs are dynamic and constantly evolve over time. In this work, we propose two Push-Relabel based algorithms for processing dynamic graphs on GPUs. The key novelty of our algorithms is their ability to efficiently handle both increments and decrements in edge capacities together when they appear in a batch. We illustrate the efficacy of our algorithms with a suite of real-world graphs. Overall, we find that for small updates, dynamic recomputation is significantly faster than a static GPU-based \maxflow. 
\end{abstract}

\maketitle

\pagestyle{\vldbpagestyle}
\begingroup\small\noindent\raggedright\textbf{PVLDB Reference Format:}\\
\vldbauthors. \vldbtitle. PVLDB, \vldbvolume(\vldbissue): \vldbpages, \vldbyear.\\
\href{https://doi.org/\vldbdoi}{doi:\vldbdoi}
\endgroup
\begingroup
\renewcommand\thefootnote{}\footnote{\noindent
This work is licensed under the Creative Commons BY-NC-ND 4.0 International License. Visit \url{https://creativecommons.org/licenses/by-nc-nd/4.0/} to view a copy of this license. For any use beyond those covered by this license, obtain permission by emailing \href{mailto:info@vldb.org}{info@vldb.org}. Copyright is held by the owner/author(s). Publication rights licensed to the VLDB Endowment. \\
\raggedright Proceedings of the VLDB Endowment, Vol. \vldbvolume, No. \vldbissue\ %
ISSN 2150-8097. \\
\href{https://doi.org/\vldbdoi}{doi:\vldbdoi} \\
}\addtocounter{footnote}{-1}\endgroup

\ifdefempty{\vldbavailabilityurl}{}{
\vspace{.3cm}
\begingroup\small\noindent\raggedright\textbf{PVLDB Artifact Availability:}\\
The source code, data, and/or other artifacts have been made available at \url{https://github.com/ShruthiKannappan/dyn_maxflow}.
\endgroup
}

\section{Introduction} \label{sec intro}
Graphs have emerged as a critical data structure for modeling relationships in a wide range of real-world applications, from social networks and recommendation systems to transportation and telecommunication infrastructures. With rapid data growth, these graphs have become massive, often containing millions of vertices and billions of edges. Processing such large-scale graphs using sequential algorithms is practically inefficient. As a result, parallelisation of graph algorithms, on CPUs, GPUs, and distributed systems, has become essential for achieving industry-scale performance.

Furthermore, many real-world graphs are \textit{dynamic} ---their structure evolves gradually over time due to the addition or deletion of nodes and edges, or updates to edge weights. Efficiently handling such changes without recomputing results from scratch presents significant algorithmic and engineering challenges, especially in parallel.
One such core problem in graph theory is the \textit{maximum flow problem} (\maxflow), which involves determining the maximum amount of flow that can be sent from a source node to a sink node in a \textit{flow network}, where each edge has a capacity constraint. Some notable use cases of \maxflow include:
\begin{itemize}
\item Network routing: optimising bandwidth allocation in communication and transportation networks. 
\item Image Segmentation: In computer vision tasks, image partitioning into foreground and background regions. 
\item Bipartite Matching: Addressing assignment problems in
job placement and resource distribution domains. 
\item Circulation and logistics: Managing flow in supply chains and transportation systems to improve efficiency.
\end{itemize}
Over the decades, numerous sequential and parallel algorithms have been proposed for computing maximum flow, including the well-known Goldberg-Tarjan algorithm~\cite{goldberg1988}, popularly known as \textit{Push Relabel}. The parallel version of this algorithm has paved the way for several implementations of \maxflow algorithms which exploit the capabilities of many-core GPUs. These efforts have led to a significant reduction in the execution time for static graphs. However, 
computing the maximum flow on \textit{dynamic graphs} remains a challenge. 

\textbf{Contributions.} This work addresses the need for efficient parallel \maxflow computation on dynamic graphs. We propose two dynamic \maxflow algorithms - the Dynamic Push Relabel Algorithm and the Dynamic Push Pull Algorithm, to run efficiently on the GPU. We also propose a static \maxflow algorithm that computes the solution on the initial graph. The resulting state serves as input to our dynamic algorithm, which incrementally updates the max-flow in response to batches of edge-capacity changes, thereby avoiding costly full re-computation. Unlike prior GPU-based approaches that are limited to static max-flow, our algorithms efficiently support dynamic updates by reusing the residual graph to maintain the solution across batches of edge changes. In contrast to CPU-based dynamic methods that rely on partial or full recomputation, our design is entirely GPU-resident: once the graph and its updates are loaded into GPU memory, all computations proceed on the GPU without CPU intervention. These features together enable indefinite maintenance of \maxflow under evolving inputs while avoiding costly CPU–GPU data transfers. The Dynamic Push Pull Algorithm works on two independent subgraphs, providing additional flexibility for parallelism.

We present two implementations of the Dynamic Push Relabel Algorithm - Topology Driven and Data-Driven. We also present the Push Pull Stream implementation for the Dynamic Push Pull Algorithm. 
The experimental evaluation of these implementations in CUDA demonstrates that the dynamic \maxflow algorithms significantly outperform traditional re-computation strategies in performance. 
Hence, demonstrating the practical effectiveness of our dynamic approach for large-scale,  continuously evolving datasets.

The rest of this paper is organised as follows:
Section~\ref{sec back} presents the problem and existing work in the domain.
Section~\ref{sec static} presents the static maxflow algorithm. Section~\ref{sec dynamic} describes the dynamic maxflow algorithms and how they reuse computations from the static algorithm.
Section~\ref{sec impl} talks about the various implementation optimisations in CUDA code.
Section~\ref{sec experiments} reports on the experimental assessment of the CUDA code.
Section~\ref{sec conclusion} concludes the paper and outlines potential directions for future work.

\section{Background and Related Work} \label{sec back}
We first formally define the \maxflow problem and then discuss the relevant related work.
\subsection{Maximum Flow Problem}
\label{sec constraints}
Given a directed graph $G = (V, E)$ where each edge $(u, v) \in E$ has capacity $c(u, v) \geq 0$, 
and two distinguished vertices: a \emph{source} $s$ and a \emph{sink} $t$, 
the objective is to determine a flow function $f : V \times V \rightarrow \mathbb{R}$ that maximizes the total flow 
from $s$ to $t$, defined as
\[
|f| = \sum_{v \in V} f(s, v) - \sum_{v \in V} f(v, s) = \sum_{v \in V} f(v, t) - \sum_{v \in V} f(t, v),
\]
subject to following conditions:
\begin{itemize}
    \item Capacity constraint: $0 \leq f(u, v) \leq c(u, v)$.
    \item Flow conservation at intermediate vertices: $\forall v \in V \setminus \{s, t\}$, $\sum_{u \in V} f(u, v) = \sum_{w \in V} f(v, w)$.
\end{itemize}

\subsection{Related Work}
The \maxflow problem has seen substantial algorithmic progress over the decades. 
The earliest approach, the augmenting path method by Ford and Fulkerson ~\cite{ford1956} finds (augmenting) paths from source to sink along which one can push more flow. This process continues until no such paths remain. Though conceptually simple, it is pseudo-polynomial in the maximum flow value. Edmonds and Karp~\cite{edmonds1972} refined this method by using BFS to choose the shortest augmenting path, leading to a time complexity of $\mathcal{O}(|V||E|^2)$. Dinitz~\cite{dinitz1970} introduced the concept of layered residual graphs and blocking flows, improving the runtime to $\mathcal{O}(|V|^2|E|)$.

A significant conceptual shift occurred with Karzanov~\cite{karzanov1974}, who introduced the idea of preflows --- allowing temporary violation of flow conservation --- to speed up convergence. Building on this, Goldberg and Tarjan proposed the \textit{Push-Relabel} algorithm~\cite{goldberg1988} which maintains a preflow and adjusts node heights to direct excess flow towards the sink. It has a worst-case time complexity of $\mathcal{O}(|V|^2|E|)$ and is now a standard due to its locality and efficiency.

Push-relabel algorithms are particularly well-suited for parallelisation since the push and the relabel operations can often be performed independently across nodes. However,  traditional CPU-parallel implementations~\cite{anderson1995} rely on locking mechanisms to manage concurrency, which were unsuitable for GPU architectures that require lock-free, massively parallel designs.
Hong~\cite {hong2008} proposed a lock-free push-relabel algorithm that leverages atomic operations, such as \emph{fetch-and-add}, to enable fine-grained parallelism without the overhead of locks. 

Although the lock-free push-relabel algorithm eliminates locking overhead and enables fine-grained parallelism, it is slow in practice because threads may perform many unnecessary push and relabel operations, leading to inefficient progress. Prior works have illustrated that one can improve the practical performance of the push-relabel algorithm by employing heuristics such as \emph{global relabeling} and \emph{gap relabeling}. The height $h$ of a vertex guides the flow direction towards the sink. The global relabeling heuristic updates vertex heights to their shortest distance (number of edges) from the sink using a backward breadth-first search (BFS) in the residual network, while the gap relabeling heuristic identifies gaps in the set of active heights and deactivates unreachable vertices.

Wu et al.~\cite{wu2012gems} provide a comprehensive overview of efficient CUDA algorithms for the maximum flow problem. They discuss the design and optimisations of parallel push-relabel algorithms on GPUs. 
Zenghyu et al \cite{he10} proposed a GPU-adapted push–relabel algorithm, addressing the difficulty of applying traditional lock-based designs on CUDA. Their hybrid CPU--GPU approach with dynamic switching 
achieves up to $2\times$ speedup over Goldberg's push--relabel implementation.
Brand et al. \cite{ brand23} present a randomized algorithm for maintaining a $\mathbf{(1-\epsilon)}$-approximate maximum flow in dynamic, incremental, capacitated graphs under edge insertions, achieving total time $O(m \sqrt{n} \cdot \epsilon^{-1})$.
A relatively recent work by Khatri et al.~\cite{Khatri2022} (referred to as \textsf{alt-pp} in this work since they perform push and pull in alternate iterations) present several optimisations and approximation techniques for maximum flow computation, including a pull-based algorithm for dynamic scenarios. Our work quantitatively compares against \textsf{alt-pp}, and demonstrates consistent performance improvements. We also complement these empirical results with a formal proof of correctness.
\vspace{-0.5em}
\subsection{Notation and Terminology}

Given a directed graph $G = (V, E)$ and a flow function $f$, the residual capacity is defined as $c_f(u,v) = c(u,v) - f(u,v) + f(v,u)$, and the residual graph as $G_f = (V, E_f)$, where $E_f = \{(u,v) \mid u,v \in V,\, c_f(u,v) > 0\}$. The excess function is $e(u) = \sum_{w \in V} f(w,u) - \sum_{v \in V} f(u,v)$, representing the net flow into vertex $u$. A vertex $u \in V \setminus \{s,t\}$ is \textit{overflowing} if $e(u) > 0$ and \textit{deficient} if $e(u) < 0$. 

An integer-valued height function $h(u)$ is defined for each $u \in V$, where $u$ is higher than $v$ if $h(u) > h(v)$. The push-valid height function $h_+$ disallows steep downward edges, i.e., $\forall (u,v) \in E_f,\; h_+(u) \leq h_+(v) + 1$.

\section{Our GPU-Static-Maxflow Algorithm}
\label{sec static} 
Existing static GPU-based \maxflow algorithms~\cite{he10, wu2012gems} typically rely on global relabeling and divide the computation into two phases:\\(i) computing the \maxflow and Min-cut, which requires $\mathcal{O}(|V|^3)$ or $\mathcal{O}(|V|^2|E|)$ time depending on the vertex processing order; and 
(ii) constructing the corresponding residual graph, which takes $\mathcal{O}(|E|\log|V|)$ time. In this paper, we focus on the first stage of the algorithm, where we can find the min-cut and the \maxflow value. 

In the existing GPU-based \maxflow algorithms, termination is usually determined by a global variable $ExcessTotal$, which tracks the total excess of active vertices. This allows Push--Relabel iterations to continue until no vertex can push flow to the sink. $ExcessTotal$ is initialized as the total flow out of the source. It is later updated when vertices are marked as unreachable during global relabeling ($ExcessTotal \gets ExcessTotal-e (u)$ for each such $u$). Importantly, an update of $ExcessTotal$ requires a sequential for-loop check over all vertices together with conditional marking of vertices that cannot reach the sink. These operations are carried out on the CPU, since maintaining and updating a single global counter with conditional checks is not well-suited for massively parallel GPU execution. As a result, both the computation of $ExcessTotal$ and global relabeling incur frequent and costly memory transfers between CPU and GPU. 


Our static GPU \maxflow algorithm eliminates this bottleneck by performing global relabeling entirely on the GPU. The termination is dictated by checking if an active vertex exists .
This avoids costly host--device communication. We present our static \maxflow algorithm as Algorithm~\ref{static_mf}. The first \texttt{for} loop (Line~\ref{line:firstfor}) initializes the height ($h_+$)  and excess ($e$) for every vertex to be 0.
The second \texttt{for} loop (Line~\ref{line:secondfor}) initializes the residual capacity of every edge $c_f$ to be its capacity as there is no flow through them. All these initialisations are fully parallel and are executed on the GPU.
The third \texttt{for} loop (Line~\ref{line:thirdfor}) sends capacity amount of flow through every edge leaving the source. This could be made parallel by ensuring the $e$ is updated using atomics. 

The algorithm then enters the while loop that continues till there are active vertices that can push flow to the sink. The iteration first initializes a vector of active vertices $A$ which have a positive excess ($e(v) > 0$) and whose height is less than $|V|$. 
Vertices that cannot reach the sink are assigned a height of $|V|$, making them inactive. A backward BFS is then called on height($h_+$). Next, the \texttt{PushRelabel} kernel is launched.
 In Algorithm~\ref{pr_kernel}, each active vertex performs a \emph{push} or a \emph{relabel} operation, up to \texttt{KERNELCYCLES} iterations.

\begin{algorithm}[H]
\caption{GPU-Static-Maxflow}\label{static_mf}
\begin{algorithmic}[1]  
\ForAll{$u\in V$}   \label{line:firstfor} \Comment{in parallel}
\State $e(u) \gets 0$
\State $h_+(u) \gets 0$ 
\EndFor
\ForAll{$(u, v) \in E$}    \label{line:secondfor} \Comment{in parallel}
    \State $c_f(u, v) \gets c(u,v)$
    \State $c_f(v, u) \gets c(v,u)$
\EndFor
\ForAll{$(s, u) \in E$}    \label{line:thirdfor} \Comment{saturate outward edges of source}
    \State $c_f(s, u) \gets 0$
    \State $c_f(u, s) \gets c(u,s) + c(s,u)$
    \State $e(u) \gets c(s,u)$
    \State $e(s) \gets e(s) - c(s,u)$
\EndFor
\While{$A\neq \phi$} \label{line:while} \Comment{$A=\{u\in V-\{s,t\}| e(u)>0 \& h_+(u)<|V| \}$}
\State $h_+(u)\gets |V| $ , $h_+(t)\gets 0$ $\forall u \in V \textbackslash\{s\}$  
    \State BFS\_Backward() 
    \State PushRelabel() \Comment{Algorithm~\ref{pr_kernel}}
    \State RemoveInvalidEdges() \Comment{Algorithm~\ref{invalid_edges_kernel}}
\EndWhile
\State $maxflow\gets e(t)$ \label{line:last}
\end{algorithmic}
\end{algorithm}

\begin{algorithm} [H]
\caption{\texttt{PushRelabel} on a vertex u  }\label{pr_kernel}
\begin{algorithmic}[1]  
\State $cnt\gets 0$
\While{$cnt<KERNEL CYCLES$}
\State $cnt$++
    \If{$h_+(u)<|V|$ $ \&\& e(u)>0  $}
    \State $e' \gets e(u)$
    \State $\hat{v} \gets \textit{null}$
    \State $\hat{h} \gets \infty$
    \ForAll{ $c_f(u, v) > 0$}
        \State $h' \gets h_+(v)$
        \If{$h' <\hat{h}$}  \label{line:hhat}
            \State $\hat{v} \gets v$
            \State $\hat{h} \gets h'$
        \EndIf
    \EndFor \Comment{$\hat{v}$ is $u$'s lowest neighbor in $E_f$}
    \If{$h_+(u) > \hat{h}$} \Comment{push$(u, \hat{v})$ is applicable}
        \State $d \gets \min(e', c_f(u, \hat{v}))$
        \State $c_f(u, \hat{v}) \gets c_f(u, \hat{v}) - d$ \Comment{Atomic fetch and subtract}
        \State $c_f(\hat{v}, u) \gets c_f(\hat{v}, u) + d$ \Comment{Atomic fetch and add}
        \State $e(u) \gets e(u) - d$ \Comment{Atomic fetch and subtract}
        \State $e(\hat{v}) \gets e(\hat{v}) + d$ \Comment{Atomic fetch and add}
    \Else \Comment{lift$(u)$ is applicable}
        \State $h_+(u) \gets \hat{h} + 1$
    \EndIf
    \EndIf
\EndWhile
\end{algorithmic}
\end{algorithm} 
 During the \emph{push} step (Lines~16--19), the residual capacities~($c_f$) and excess values~($e$) are updated using atomic operations to ensure correctness under parallel execution. All remaining steps do not require any synchronisation primitives.

After the \texttt{PushRelabel} kernel, concurrent updates may temporarily violate 
the height invariant of the serial push algorithm. An edge $(u,v)$ is considered \emph{invalid} if it violates the condition 
$h_+(u) \leq h_+(v) + 1$.
Such violations can occur when a thread processing vertex~$u$ performs a 
\texttt{relabel} operation while another thread simultaneously pushes the flow 
toward~$u$ using an outdated height value. 
This introduces a steep downward edge in the residual graph. \texttt{RemoveInvalidEdges} kernel~\ref{invalid_edges_kernel} detects and removes these invalid edges in parallel by saturating the edge in the residual graph.  Each thread iterates over the outgoing edges of its assigned vertex. Since the residual capacity ($c_f$) of each edge is dealt by the source only, 
no synchronization is required. However, the excess function $e$ could be updated by more than 1 thread; hence atomics are used. The backward BFS, \texttt{PushRelabel}, and \texttt{RemoveInvalidEdges} kernels are invoked in a loop that terminates once no active vertices remain.
After Algorithm~\ref{static_mf} terminates, the preflow contains no active vertices, and the excess accumulated at the sink corresponds to the maximum flow (Line~\ref{line:last}).
\begin{algorithm}[H]
\caption{     \texttt{RemoveInvalidEdges} on  a vertex u    ($h_+$ invalid)}\label{invalid_edges_kernel}

\begin{algorithmic}[1]  

\For{$(u,v)\in E_f$}
 \If{$h_+(u)>h_+(v)+1$}
        \State $e(u) \gets e(u) - c_f(u,v)$ \Comment{Atomic fetch and subtract}
        \State $e(v) \gets e(v) + c_f(u,v)$ \Comment{Atomic fetch and add}
        \State $c_f(v,u) \gets c_f(v,u) + c_f(u,v)$ 
        \State $c_f(u,v) \gets 0$
\EndIf
\EndFor
\end{algorithmic}
\end{algorithm}

\subsection{Proof of Correctness}
\lemma{At the end of each iteration of Algorithm~\ref{static_mf}, the push-valid height function $h_+(v)\leq d(v)$, where $d(v)$ represents the shortest BFS distance (in terms of the number of edges) from $v$ to sink($t$).}\label{distheight}
\proof After the push-relabel kernel completes, the parallel algorithm might violate the height invariant of the serial algorithm, that is, $\forall (u,v)\in E_f$, $h_+(u)\leq h_+(v)+1$. This situation arises when a vertex $u$ performs a relabel, while $v$ might have read a stale value of $h_+(u)$ before $u$ performed a relabel, and hence the edge $(u,v)$ could be added as a result of a push from $v$ to $u$. However, the relabeling of $u$ makes the newly added edge a steep downward edge. The remove-invalid edges (Algorithm~\ref{invalid_edges_kernel}) restores this invariant by removing such invalid edges. 
Hence at the end of each iteration, $\forall (u,v)\in E_f$, $h_+(u)\leq h_+(v)+1$. 

Let us assume $\exists v$ such that  $d(v)<h_+(v)$ for some vertex $v$. This means that there exists a steep downward edge, which is not possible by the invariant. Hence, the assumption is false, and we conclude that for every vertex \( v \), the condition \(h_+ (v) \leq d(v) \) holds. 

\theorem{Height of every vertex is non-decreasing. }\label{heightinc}
\proof The height of any vertex is changed by the following two operations:
(i) relabel, which increases the height of the vertex by at least 1, and (ii) backward BFS, which never decreases the height, because it assigns $d(v)$ to $h_+(v)$ (from Lemma~\ref{distheight}). 

\theorem{After the termination of the Algorithm~\ref{static_mf}, an  $S, T$ cut can be obtained whose forward edges are saturated by the preflow and the backward edges have no flow in them with the cut capacity $C(S, T)$ = $e(t)$}\label{abcut}.
\proof 
After the algorithm terminates, the vertices of the graph can be partitioned into two disjoint sets $S$ and $T$, where \\$S = \{u \mid u \in V \ \text{and}\ h(u)=|V|\}$ and 
$T = \{v \mid v \in V \ \text{and}\ h(v)<|V|\}$. 
Since the sink $t$ is assigned a height of $0$, it belongs to $T$, and since all the edges outgoing from the source $s$ are saturated, $s$ cannot reach any vertex and thus lies in $S$.

For every edge $(a,b) \in E$ with $a \in S$ and $b \in T$, the residual capacity $c_f(a,b)=0$. Otherwise, if $c_f(a,b) > 0$, then $h(a) < |V|$, which contradicts $a \in S$. Therefore, $f(a,b)=c(a,b)$. Similarly, for every edge $(b,a)\in E$ with $a\in S$ and $b\in T$, we must have $f(b,a)=0$, because if $f(b,a)>0$ then $c_f(a,b)>0$, implying $h(a)<|V|$, which again contradicts $a\in S$. Hence, $(S,T)$ forms an $s$–$t$ cut whose forward edges are saturated and whose backward edges carry no flow.

Let $e(T)=\sum_{u\in T} e(u)$. Since there are no active vertices upon termination, we have $e(T)=e(t)$. 
By the definition of excess,
\[
e(T) = \sum_{u\in T} e(u)
      = \sum_{u\in T}\sum_{v\in V} f(v,u) 
        - \sum_{u\in T}\sum_{v\in V} f(u,v)
\]
This can be rewritten as
\[
e(T)
 = \sum_{u\in T}\sum_{v\in S} f(v,u) 
   - \sum_{u\in T}\sum_{v\in S} f(u,v),
\]
because internal flows within $T$ cancel out. From the argument above, for all $u\in T$ and $v\in S$, we have $f(u,v)=0$ and $f(v,u)=c(v,u)$. 
Therefore, 
\[
e(T) 
=  \sum_{u\in T}\sum_{v\in S} c(v,u)
= C(S,T) = e(t),
\]
which shows that the excess at the sink equals the capacity of the $s$–$t$ cut, C(S,T). \qed

\lemma{ Upon termination of the algorithm $\forall v\in V-\{s,t\}$ if $e(v)>0$ then there exists a path from v to s in the residual graph.} \label{path1}
\proof 
Upon termination, there are no vertices with $e(v) > 0$ and $h(v) < |V|$. 
Hence, if there exists any $v \in V \setminus \{s,t\}$ such that $e(v) > 0$, then $h(v)=|V|$ and therefore $v \in S$.

Assume, for contradiction, that there exists a vertex $u_0 \in S$ with $e(u_0) > 0$ that has no path to $s$. We
define the sets 
$X = \{u \in S \mid u \not\rightsquigarrow s\}$ and 
$Y = \{u \in S \mid u \rightsquigarrow s\}$.
By assumption, $u_0 \in X$. 

Since this is a preflow, $e(v)\ge 0$ for all $v \in V\setminus\{s,t\}$. 
Consider 
\[
e(X)=\sum_{u\in X} e(u)
=\sum_{u\in X}\sum_{v\in V} f(v,u)-\sum_{u\in X}\sum_{v\in V} f(u,v)
\]

This can be rewritten as
\[
\sum_{u\in X}\sum_{v\in Y}f(v,u)+\sum_{u\in X}\sum_{v\in T}f(v,u)
 -\sum_{u\in X}\sum_{v\in Y}f(u,v)-\sum_{u\in X}\sum_{v\in T}f(u,v)
\]

From Theorem~\ref{abcut}, for all $v\in T$ and $u\in X$, $f(v,u)=0$. 
Thus, $\sum_{u\in X}\sum_{v\in T}f(v,u)=0$. Since $u_0\in X$, $e(X)>0$. 
Therefore, there exists some $u\in X$ and $v\in Y$ such that $f(v,u)>0$. 
This implies $(u,v)\in E_f$ and hence $u\rightarrow v \rightsquigarrow s$, contradicting the assumption that $u\not\rightsquigarrow s$.
\qed

\theorem{At the algorithm termination, e(t) is the maxflow.}
\proof 
After the algorithm terminates, we obtain a preflow $f$ and a cut $(S,T)$. 
From $f$, we construct a valid flow $f^*$ by routing the excess from each overflowing vertex to the source along any augmenting path, whose existence is guaranteed by Lemma~\ref{path1}. 
These paths lie entirely within $S$ and therefore do not alter the flow across the cut $(S,T)$. 
As a result, $f^*$ is a valid flow satisfying $e(v)=0$ for all $v\in V\setminus\{s,t\}$ and $-e(s)=e(t)=|f^*|$.

By Theorem~\ref{abcut}, we have $C(S,T)=e(t)=|f^*|$. 
By the Max-Flow Min-Cut Theorem~\cite{ford1956}, $|f^*|=e(t)$ equals the value of the maximum flow in the graph.
\qed

\theorem{Algorithm~\ref{static_mf} terminates after at most $\mathcal{O}(|V|^2|E|)$ push and relabel operations.} \label{termination}

\proof 
By Lemma~\ref{heightinc}, each vertex can be relabeled at most $|V|$ times (including both local and global relabels). 
Hence, at most $|V|^2$ relabel operations are performed in total.

As in the sequential push-relabel analysis, a saturated push on an edge can only occur again after a relabel. 
Since each vertex can be relabeled at most $|V|$ times, the total number of saturated pushes is at most $|V||E|$. 
Furthermore, the number of unsaturated pushes is bounded by $4|V|^2(|V|+|E|)$.

Because the total number of push and relabel operations is finite, the algorithm eventually eliminates all active vertices and terminates. 
Global relabeling is performed only once within an iteration, which continues till push or relabel operations occur and thus is also bounded.
\qed

\textbf{Notes}
\begin{enumerate}
    \item The cut $S$ = $\{u\in V| h(u)=|V|\}$ and $T$ = $\{u\in V| h(u)<|V|\}$ is a min-cut, hence it could be used as a certificate for the output of \maxflow. 
\item In our algorithm, we can indirectly control the number of times global relabel is called by tuning the \textit{KERNELCYCLES} parameter, which is the maximum number of pushes/relabels a kernel executes for an active vertex. A very low value for this parameter may lead to excessive global relabel calls, resulting in increased computational overhead and decreased performance. Conversely, a high value could lead to vertices pushing flow in the wrong direction. 
\item Line~\ref{line:hhat} in Algorithm~\ref{pr_kernel} is different from the lock-free algorithm where $\hat{h}$ is used instead of $h(\hat{v})$ to ensure that the height in one execution is read at most once in order to prove Lemma 3 in Hong~\cite{hong2008}.

\end{enumerate}
\section{Dynamic Maxflow} \label{sec dynamic}

The primary objective of this work is to develop dynamic algorithms for the \maxflow problem. When a graph is updated through a batch of edges with modified capacities, a dynamic \maxflow algorithm \textit{adjusts} the maximum flow without rerunning the algorithm from scratch on the updated graph. To ensure enough work for the GPU, we expect that the updates are batched. Each batch is a set of edges whose new capacities may be either higher or lower than their previous values. The goal, therefore, is to design an efficient approach that minimises computational time while accurately recomputing the maximum flow after each batch update. We present two algorithms. The first one is conceptually similar to the GPU static \maxflow presented in Section ~\ref{sec static}, but for dynamic updates and in parallel. The second assumes the $S,T$ min-cut remains unchanged, performing simultaneous push--relabel on $T$ and pull on $S$, followed by a selective push--relabel to correct the Min-cut differences.

\subsection{Dynamic Push Relabel Algorithm }
The dynamic processing continues from the state computed by the static \maxflow (Algorithm~\ref{static_mf}). Two cases arise depending upon the edge-capacity increase and decrease. When the edge-capacity increases, there is a possibility that \maxflow increases. However, this happens if the edge is part of the min-cut. On the other hand, when the edge-capacity decreases, the current flow may reduce. However, this can happen whether the edge is part of the min-cut or not. 
In an extreme case, the edge capacity can become zero -- indicating removal of that edge. Similarly, a jump from zero capacity to a positive capacity simulates addition of an edge.

Our Dynamic Push-Relabel is presented in  Algorithm~\ref{dynamic_mf}. It begins with processing of the input updates, which is presented in Algorithm~\ref{updates_processing}.  Algorithm~\ref {updates_processing} (Lines 1--3) updates the residual capacity of each edge in parallel by updating the edge capacity while maintaining the flow. However the residual capacity can become negative if the new capacity is lower than the original flow. 
Hence, it reduces the flow to the capacity of that edge by sending back the extra flow (Lines 4--11).  Since the flow through some edges may have been reduced (potentially creating deficient vertices), the excess of every vertex must be recalculated. 

Similar to the static algorithm, the source saturates all forward edges from itself.
The algorithm loops through an iteration of  Backward BFS, Push-Relabel, and Remove-Invalid edges till there are no more $active$ vertices. The definition of $active$ vertex is the same, where excess is positive and the height is $<|V|$.\begin{algorithm} [H]
\caption{Dynamic Push Relabel}\label{dynamic_mf}
\begin{algorithmic}[1]  
\State Call Updates Processing \Comment{Algorithm~\ref{updates_processing}}
\State Calculate excess for each vertex in parallel
\ForAll{$(s, u) \in E$} 
    \State $c_f(s, u) \gets 0$
    \State $c_f(u, s) \gets c(u,s) + c(s,u)$
    \State $e(u) \gets c(s,u)$
    \State $e(s) \gets e(s) - c(s,u)$
\EndFor
\ForAll{$u \in V$}\label{line:dynpr-start} \Comment{in parallel}
\State  $h_+(u)\gets0$  
\EndFor

\While{$A\neq \phi$} \Comment{$A=\{u\in V-\{s,t\}| e(u)>0 \& h_+(u)<|V| \}$}
\For{$u \in V $} \Comment{in parallel}
    \If{$u==t$ or $(e(u)<0$ and   $u\neq s)$}
        \State $h_+(u) \gets 0$
    \Else
    \State $h_+(u) \gets |V|$
    \EndIf
\EndFor
    \State BFS\_Backward() 
    \State PushRelabel() \Comment{Algorithm~\ref{pr_kernel}}
    \State RemoveInvalidEdges() \Comment{Algorithm~\ref{invalid_edges_kernel}}
\EndWhile   \label{line:dynpr-end} 
\State $maxflow \gets 0$
\ForAll{$u \in V$}
    \If{$h(v)==0$}
        \State $maxflow \gets maxflow + e(v)$
    \EndIf
\EndFor
\end{algorithmic}
\end{algorithm} 

The key difference from static processing is that deficient vertices are now effectively treated as {\it sinks}.
 Hence, all deficient vertices along with the sink are assigned height 0, and backward BFS is carried out from this state. The active vertices do the same task, but now the flow is directed either to the sink or to any of the deficient vertices. At the end, the source and the overflowing vertices will be at height $|V|$,

\begin{algorithm}[H]
\caption{Updates Processing}\label{updates_processing}
\begin{algorithmic}[1]  
\ForAll{$(u, v) \in Updates$} \Comment{in parallel}
    \State $c_f(u,v) \gets c_f(u,v)+c'(u,v)-c(u,v)$
\EndFor
\ForAll{$u \in V$} \Comment{in parallel}
    \For{$(u,v) \in E$}
    \If{$c_f(u,v)<0$} \Comment{if flow more than capacity}
        \State $c_f(v,u) \gets c_f(v,u)+c_f(u,v)$
        \State $c_f(u,v) \gets 0$
    \EndIf
    \EndFor
\EndFor
\end{algorithmic}
\end{algorithm}while the sink and the deficient vertices will be at height 0;  vertices with height $>0$ and $<|V|$, the excess would be 0. The final \maxflow value is computed by simply taking the sum of the excess of vertices at height 0.

\subsubsection{Proof of correctness}
\lemma{At the end of the algorithm , $\forall v\in V-\{s,t\}$ if $e(v)<0,$ then there exists an augmenting path from t to v in the residual graph}\label{path2}
\proof 
Upon termination, the algorithm ensures that no active vertices remain, i.e., $\forall v\in V$, if $e(v)>0$ then $h(v)=|V|$, and if $e(v)\le 0$ then $h(v)=0$.

Let $S = \{u \in V \mid h(u)=|V|\}$ and $T = \{u \in V \mid h(u)<|V|\}$. Then, for all $v \in V\setminus\{s,t\}$, if $e(v)<0$ then $h(v)=0$, implying $v \in T$.  

Assume, for contradiction, that there exists $u_0 \in T$ such that $e(u_0)<0$ and there is no path from $t$ to $u_0$. 

Define  
$X'=\{u\in T \mid t \not\rightsquigarrow u\}$ and $Y'=\{u\in T \mid t \rightsquigarrow u\}$.  
Clearly, $X'\cap Y'=\varnothing$, $X'\cup Y'=T$, and $u_0 \in X'$.  Let the total excess of $X'$:
\[
e(X') = \sum_{u\in X'} e(u) 
      = \sum_{u\in X'}\sum_{v\in V} f(v,u) - \sum_{u\in X'}\sum_{v\in V} f(u,v).
\]

Splitting by vertex sets simplifies to 
\[
\sum_{u\in X'}\sum_{v\in Y'} f(v,u) 
+ \sum_{u\in X'}\sum_{v\in S} f(v,u) 
-\sum_{u\in X'}\sum_{v\in Y'} f(u,v)
-\sum_{u\in X'}\sum_{v\in S} f(u,v).
\]

Using Theorem~\ref{abcut}, we can state that $\forall u\in X' \subset T$ and $v\in S$, $f(u,v)=0$ and $f(v,u)=c(v,u)$, we get
\[
e(X') = \sum_{u\in X'}\sum_{v\in Y'} f(v,u) 
+ \sum_{u\in X'}\sum_{v\in S}  c(v,u) 
        - \sum_{u\in X'}\sum_{v\in Y'} f(u,v).
\]

Since $\forall u\in X', e(u)\leq0$ and  $u_0\in X'$, we have $e(X')<0$. This implies $\exists\, u\in X'$ and $v\in Y'$ such that $f(u,v)>0$, $\implies$ $(v,u)\in E_f$ and thus there is a residual path $t\rightsquigarrow v\rightarrow u$. This contradicts the assumption that $t$ cannot reach $u$. 
\qed
\theorem{After the algorithm terminates, the computed \\maxflow is the maximum flow in the updated graph.}
\proof At the end of the algorithm, we have a Min-cut $S, T$ ( $S = \{u \in V \mid h(u)=|V|\}$ and $T = \{u \in V \mid h(u)<|V|\}$). 
From $f$, we can construct $f_1$ by sending the excess flow from overflowing vertices through any augmenting path
to the source, which is guaranteed to exist from Lemma~\ref{path1}. 
Now $f_1$ is left with deficient vertices. We can now send flow from t to all such v through the augmenting paths, which are guaranteed to exist from Lemma~\ref{path2}. On continuing this process till there no deficient vertices, we can obtain a valid flow $f^*$. Note that these paths lie within S and T, respectively, hence the cut (S, T) is not disturbed. Similar to theorem \ref{abcut}, we can show that $C(S, T)$ = $|f^*|$ = $e'(t)$ ($e'$ is excess function of $f^*$). By the \maxflow Min-cut theorem, $f^*$ is a maximum flow for the graph, and S, T is a min-cut. The flow sent from $t$ to the deficient vertices, causes $e'(t)$ = $\sum_{v\in V-\{s,t\} | e(v)<0 }e(v)+e(t)$, which is same as  $\sum_{v| h_+(v)==0 }e(v)$ = $maxflow$.
\\\theorem{Height of every vertex is non-decreasing.}\label{heightinc2}

\proof 
As the algorithm proceeds, some deficient vertices might become overflowing or stable ($e(v)\geq0$); in such cases, other deficient vertices and the sink will guide the flow. Note that the d(v) for a vertex v does not decrease because of this conversion from \textit{deficient} to \textit{non-deficient}, since d(v) represents the minimum from distances to the sink and deficient vertices. Hence, d(v) is increasing in nature. The rest of the proof is similar to Theorem \ref{heightinc}
\\\textbf{Note}: The proof of termination is similar to that of the static algorithm (Theorem~\ref{termination}) by taking Theorem \ref{heightinc2} into consideration. 

\subsection{Dynamic Push Pull Algorithm}
Similar to push relabel, where overflowing vertices push flow to neighbours, one can think of pull relabel as processing of deficient vertices, i.e $e(u)<0$ instead. Deficient vertices can pull flow from neighbours by increasing the flow in incoming edges. However, this requires a different height function $h_-$, where steep upward edges are not allowed ($\forall (u,v)\in E_f, h_-(v)\leq h_-(u)+1$). In contrast to push relabel, $h_-(s)=0$ and $h_-(t)=|V|$. The deficient vertices pull flow from a lower neighbor if one exists; otherwise, they relabel. Hence, Pull Relabel is analogous to push relabel, as is its proof.

\begin{algorithm}[H]
\caption{\texttt{PullRelabel} on a vertex u  }\label{pull_kernel}
\begin{algorithmic}[1]  
\State $cnt\gets 0$
\While{$cnt<KERNEL CYCLES$}
    \State $cnt++$
    \If{$h_-(u)<|V|$ $ \&\& e(u)<0  $}
    \State $e' \gets e(u)$
    \State $\hat{v} \gets \textit{null}$
    \State $\hat{h} \gets \infty$
    \ForAll{$c_f(v, u) > 0$  }
        \State $h' \gets h_-(v)$
        \If{$h' <\hat{h}$}
            \State $\hat{v} \gets v$
            \State $\hat{h} \gets h'$
        \EndIf
    \EndFor \Comment{$\hat{v}$ is $u$'s lowest neighbor in $E_f$}
    \If{$h_-(u) > \hat{h}$} \Comment{pull flow from $\hat{v}$ to $u$ is applicable}
        \State $d \gets \min(e', c_f(\hat{v},u))$
        \State $c_f(\hat{v},u) \gets c_f(\hat{v},u) - d$ \Comment{Atomic fetch and subtract}
        \State $c_f(u,\hat{v}) \gets c_f(u,\hat{v}) + d$ \Comment{Atomic fetch and add}
        \State $e(u) \gets e(u) + d$ \Comment{Atomic fetch and add}
        \State $e(\hat{v}) \gets e(\hat{v}) - d$ \Comment{Atomic fetch and Subtract}
    \Else \Comment{lift$(u)$ is applicable}
        \State $h_-(u) \gets \hat{h} + 1$
    \EndIf
    \EndIf
    
\EndWhile

\end{algorithmic}
\end{algorithm}
\begin{algorithm}
\caption{\texttt{RemoveInvalidEdges} on  a vertex u    ($h_-$ invalid)}\label{invalid_pull_edges_kernel}
\begin{algorithmic}[1]  

\For{$(u,v)\in E_f$}
 \If{$h_-(v)>h_-(u)+1$}
        \State $e(u) \gets e(u) - c_f(u,v)$ \Comment{Atomic fetch and subtract}
        \State $e(v) \gets e(v) + c_f(u,v)$ \Comment{Atomic fetch and add}
        \State $c_f(v,u) \gets c_f(v,u) + c_f(u,v)$ 
        \State $c_f(u,v) \gets 0$
\EndIf
\EndFor
\end{algorithmic}
\end{algorithm}
One can also use pull relabel for the dynamic maxflow algorithm by treating the overflowing vertices as the source and assigning 0 height, performing forward BFS instead of backwards BFS. 
In fact, precisely this is done by Khatri et al.~\cite{Khatri2022}, who call push and pull alternatively for the dynamic processing. However, the algorithm was not proved for correctness (height function is no longer monotone). Similarly, Juntong et al.~\cite{juntong} create a different height function and perform push and pull simultaneously -- but not in a shared-memory setting. 
Additionally, performing push and push simultaneously or in alternate iterations in our setting is not straightforward. Adopting push and pull \textbf{simultaneously} in the parallel setting creates the following problems:
\begin{itemize}
    \item The validity of the height functions is necessary for the proof of termination. It could be possible that an edge is valid in the pull height function and invalid in the push height function, while the reverse edge is valid in the opposite. In such a case or vice versa, at least one of the height functions needs to be compromised. 
    \item If a deficient vertex and an overflowing vertex are neighbors, they might attempt to perform push and pull simultaneously, possibly leading to negative residual capacity as both push flow through the same edge. 
    \item The positive and negative excess could end up cyclically chasing each other.
\end{itemize}
On the other hand, when push and pull are invoked in \textbf{alternate iterations}, proving termination becomes challenging because the height function need not be monotone for either operation. New deficient or overflowing vertices may be created, which can change the BFS roots and  decrease certain heights. Consequently, standard arguments that rely on monotonic height increase do not apply.

\begin{figure}[H]
\centering
\captionsetup[subfigure]{justification=centering, font=small}

\begin{subfigure}{0.4\linewidth}
    \centering
    \includegraphics[width=\linewidth]{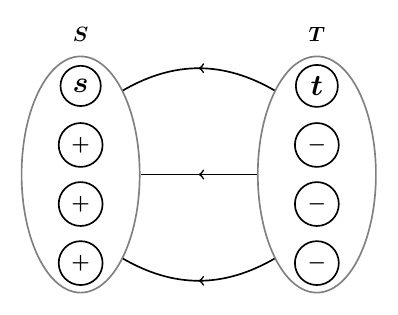}
    \caption{Static algorithm's result}
\end{subfigure}
\hfill
\begin{subfigure}{0.58\linewidth}
    \centering
    \includegraphics[width=\linewidth]{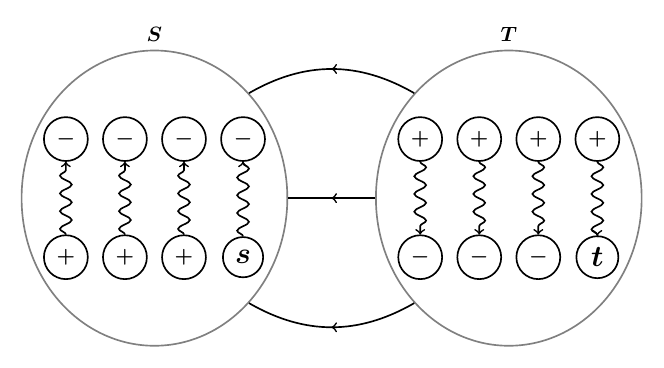}
    \caption{Retaining the cut after updates}
\end{subfigure}
\caption{Residual graph, stage 1 of Push–Pull Algorithm \ref{pp_dynamic_mf}}
\label{fig:p1p2}
\end{figure}

During the static computation, one can saturate the incoming edges to the sink and treat the resulting deficient vertices as secondary sinks, analogous to the dynamic algorithm. 
We call this approach as \textit{static push-pull algorithm}. 
Figure~\ref{fig:p1p2}(a) illustrates the resulting residual graph and the corresponding min-cut. 
The maximum flow can be obtained by summing the excess values in either partition. 
Here, $T$ represents the set of vertices that have a residual path to the sink~($t$) or any deficient vertex, 
while $S$ comprises the remaining vertices—specifically, the overflowing vertices, the source~($s$), 
and those that cannot reach any vertex in ~$S$. 
By the definition of the Min-cut, there are no residual edges  from $S$ to $T$; hence, the figure depicts only the edges from $T$ to $S$, reflecting the final residual graph after the algorithm converges.
\textbf{Algorithm:}
The Dynamic Push Pull Algorithm~\ref{pp_dynamic_mf} is inspired by the pull-relabel algorithm and assumes minor min-cut changes on updates.
\begin {algorithm}[H]
\caption{Push Pull Dynamic Maxflow}\label{pp_dynamic_mf}
\begin{algorithmic}[1]  
\For{$u\in V$}
\If{$h_+(u)<|V|$} \Comment{$h_+$ from static Algorithm}
\State $part(u)\gets T$
\Else
\State $part(u)\gets S$
\EndIf
\State $h_-(u)\gets 0, h_+(u)\gets 0$
\EndFor
\State Call Updates Processing \ref{updates_processing}
\ForAll{$u\in S$ and $v\in T$ and $c_f(u,v)>0$}  
\State $c_f(v,u) \gets c_f(v,u)+c_f(u,v)$ \Comment{remove S to T edges}
\State $c_f(u,v) \gets 0$ 
\EndFor
\State calculate excess for each vertex in parallel
\While{$A\neq \phi$} \Comment{$A=\{u\in V-\{s,t\}| $ $(e(u)>0 $ and $ h_+(u)<|V| )$ or $(e(u)<0 $ and $ h_-(u)<|V| )$ $\}$}
\For{$u \in V $} \Comment{in parallel}
    \State $h_+(u) \gets |V|$, $h_-(u) \gets |V|$
    \If{$u==t$ or $(e(u)<0$ and   $part(u)==T)$}
        \State $h_+(u) \gets 0$
    \EndIf
    \If{$u==s$ or $(e(u)>0$ and   $part(u)==S)$}
        \State $h_-(u) \gets 0$
    \EndIf
\EndFor
    \State Perform Forward BFS for $h_-$ and Backward BFS for $h_+$.
    \State Launch push (Algorithm~\ref{pr_kernel}) and pull (Algorithm~\ref{pull_kernel})  kernels in parallel
    \State Launch \texttt{RemoveInvalidEdges}  kernel for $h_+$ (Algorithm~\ref{invalid_edges_kernel}) after push kernel and $h_-$ (Algorithm~\ref{invalid_pull_edges_kernel}) after pull kernel
\EndWhile
\For{$u \in V$}
\If{$h_+(u)==|V|$ and $h_-(u)==|V|$}
    \State $part(u)\gets P$
\EndIf
\EndFor

\State Execute Lines~\ref{line:dynpr-start}--\ref{line:dynpr-end} of  Dynamic Push Relabel on subgraph $P$ 
\ForAll{$u\in V$}
\If{$part(u)==P$}
\If{$ h_+(u)<|V|$} \Comment{$h_+$ is overwritten by Dynamic Push  Relabel on $P$}
\State  $part(u)\gets T'$\Comment{$u\in B''$}
\Else 
\State $part(u)\gets S'$ \Comment{$u\in A''$}
\EndIf
\EndIf
\If{$part(u)==T$}
\State  $part(u)\gets T'$ \Comment{$u\in B$}
\EndIf
\If{$part(u)==S$}
\State  $part(u)\gets S'$\Comment{$u\in A'$}
\EndIf
\EndFor
\State $maxflow \gets 0$
\ForAll{$u \in V$}
    \If{$part(u)==T'$}
        \State $maxflow \gets maxflow + e(u)$
    \EndIf
\EndFor
\end{algorithmic}
\end{algorithm}
We initially assume that the Min-cut does not change after the updates. Based on this assumption, we saturate all edges across the original Min-cut (Lines 10--13). 
The previous cut $(S, T)$, which originally had overflowing vertices only in $S$ and deficient vertices only in $T$, may now have both overflowing and deficient vertices on either side.
However, since all edges from $S$ to $T$ are saturated, no residual path exists from any vertex in $S$ to any vertex in $T$. The resulting residual graph thus takes the form shown in Figure~\ref{fig:p1p2}(b).

Hence, overflowing vertices and the source in $S$ are assigned height $h_- = 0$ for the pull algorithm, while deficient vertices and the sink in $T$ are assigned height $h_+ = 0$ for the push algorithm (Lines 16--24). Performing Forward BFS for pull and Backward BFS for push, from the respective bases,  would ensure that the two algorithms run on disjoint sets of vertices and edges (Line 25). One can thus execute the two algorithms independently as they access disjoint sets of vertices and edges. We can thus launch \texttt{PushRelabel} and \texttt{PullRelabel} algorithms concurrently by using different streams. \texttt{RemoveInvalidEdges} for $h_+$ and $h_-$ are also independent of each other and can be launched in the corresponding streams to ensure they are launched after \texttt{PushRelabel} and \texttt{PullRelabel} respectively.  This loop continues until there are no active vertices for either of the algorithms (Lines 15--28). After this, we obtain a cut $ A', B'$ in S and a cut $A, B$  in T. 

\begin{figure}[H]
\centering
\captionsetup[subfigure]{justification=centering, font=small}

\begin{subfigure}{0.54\linewidth}
    \centering
    \includegraphics[width=\linewidth]{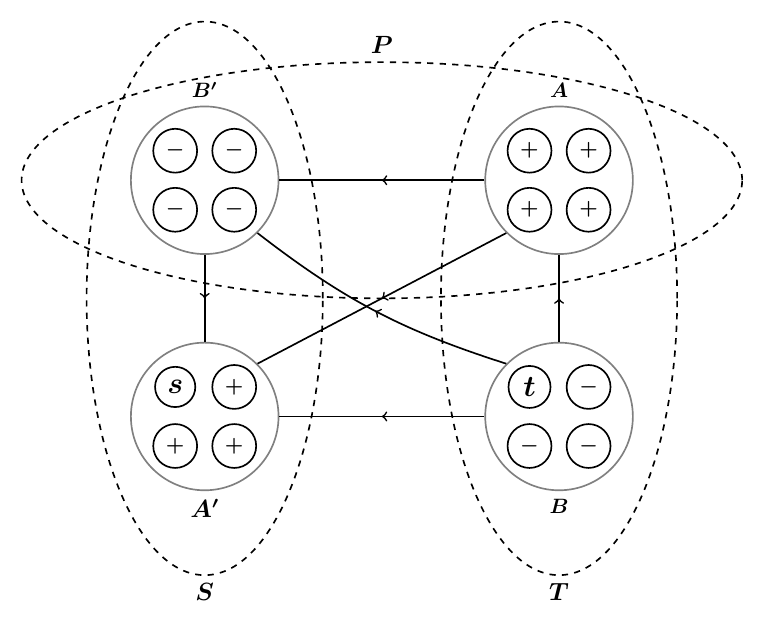}
    \caption{Post Push on T and Pull on S }
\end{subfigure}
\hfill
\begin{subfigure}{0.45\linewidth}
    \centering
    \includegraphics[width=\linewidth]{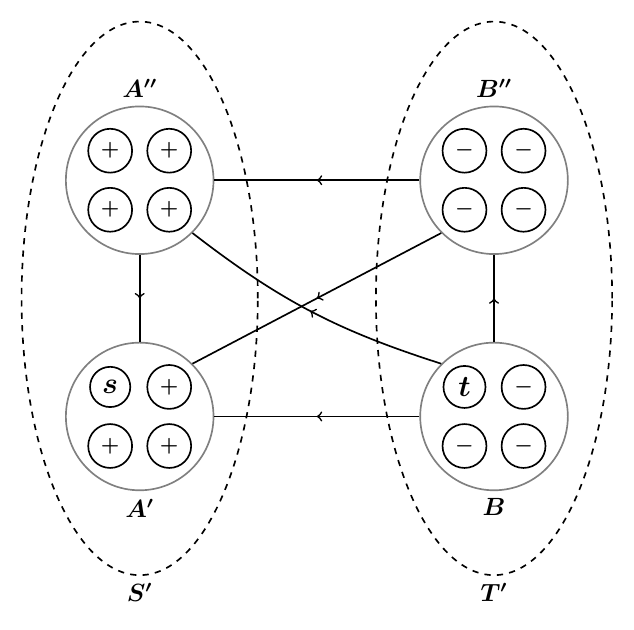}
    \caption{Final Result}
\end{subfigure}

\caption{Residual graph, stage 2 of Push–Pull Algorithm \ref{pp_dynamic_mf}}
\label{fig:p3p4}
\end{figure}

As shown in Figure~\ref{fig:p3p4}(a), the only possible path from an overflowing vertex to a deficient vertex can exist from an overflowing vertex in $A \subseteq T$ to a deficient vertex in $B' \subseteq S$. To reduce the overhead of BFS and related operations, we therefore launch the push algorithm only on this localized section $ P =  B' \cup A$ (Lines 29--34).
This process gives a cut $(A'', B'')$. We can thus construct the new min cut $(S', T')$, where $S' = A'' \cup A' $ and $ T' = B \cup B'' $ (Lines 31--41). The maxflow is thus the sum of the excess in $T'$.  

The correctness of this construction follows directly: there are no residual edges from $S'$ to $T'$, and all excess can be pushed from overflowing vertices to the source and from the sink to the deficient vertices to obtain the maxflow residual graph.
\section{Implementations and Optimisations}
\label{sec impl}
We describe the graph representation below, followed by the three dynamic \maxflow implementations: \textsf{topology-driven}, \textsf{data-driven}, and \textsf{push-pull stream}.

\subsection{Graph Representation}
Efficient neighbour access is critical for GPU-based graph algorithms. The Compressed Sparse Row (CSR) format is known for its compactness, using two arrays: offset and edgelist. However, since CSR encodes only the outgoing edges, accessing the reverse edges during push-relabel operations is non-trivial. To avoid additional lookups or pointers to reverse edges, we adopt the Bi-Directional CSR (Bi-CSR) graph representation, which explicitly stores both incoming and outgoing edges. We use zero-capacity entries to address any missing reverse edges. Each push operation modifies the forward and the reverse edges; hence, we must update the neighbours' edge list to ensure consistency. An additional array is maintained to access the reverse edge in the neighbour's edgelist, which stores the offset of the reverse edges in the CSR. The above data structures allow us to  perform push on  an edge in a constant number of memory accesses.


\subsection{Topology-Driven Processing}
In topology-driven implementations, all vertices are treated as active, and the operator is applied uniformly across the entire graph, even if some vertices have no work to perform~\cite{6569834}. This approach is straightforward to implement and does not require any additional memory. In our algorithm, the topology-driven approach launches the Push-Relabel kernel (Algorithm~\ref{pr_kernel}) and the Removing Invalid Edges kernel (Algorithm~\ref{invalid_edges_kernel}) with a configuration where the total number of threads equals the number of vertices in the graph. However, the threads for non-active vertices will be idle. 

An additional advantage of this approach is its ability to immediately process any new active vertices that become active as a result of pushes from their neighbours. However, this advantage is 
highly dependent on the graph structure and thread scheduling.

\subsection{Data-Driven / Worklist Processing}
In a data-driven model, computation is applied selectively, targeting only the nodes where work is present. 
Our empirical evaluation suggests that the number of \textit{active} vertices is small compared to the total number of vertices in the graph. 
Hence, we use a data-driven approach wherein we create a worklist of active vertices and launch the push-relabel kernel and the remove-invalid edges kernel only for the worklist. 
In this implementation, we create a worklist using atomics on a shared pointer. Each warp calculates the number of active vertices using warp-based ballot voting in CUDA. The leader of the warp then gets the offset of the warp using atomics on the shared pointer. 

\subsection{Push Pull Stream}
In this implementation, we use CUDA streams to launch the kernels of  \texttt{PushRelabel} and \texttt{PullRelabel} concurrently. Streams ensure that the \texttt{RemoveInvalidEdges} kernel for $h_+$ is launched after the \texttt{PushRelabel}  kernel, and similarly for pull. This brings independence between the two algorithms. We also incorporate data-driven processing for the two algorithms. Forward and Backward BFS are carried out simultaneously by simply launching one kernel, which scans outward and inward edges based on the partition.   For the second half (Algorithm~\ref{pp_dynamic_mf}), where only the subgraph $P$ needs processing, we create a worklist to store vertices in $P$ so that the kernels process only these vertices for BFS and other processing.

\section{Experimental Evaluation}\label{sec experiments}
We first describe our experimental setup, followed by results for incremental, decremental, and mixed updates.

\subsection{Experimental Setup}
All our experiments were carried out on NVIDIA Tesla V100-PCIE with the following specifications:
\begin{itemize}
    \item Architecture: Volta
    \item Compute capability: 7.0 
    \item Streaming Multiprocessors (SMs): 80
    \item Total Memory: 32 GB
    \item Memory Bandwidth: 900 GB/s.
    \item Max. threads per SM: 2048 
    \item Max. threads per block: 1024
\end{itemize}
We use CUDA 11.4. The CPU hosts an Intel Xeon Gold 6248 CPU with 40 hardware threads spread over two sockets, 2.50 GHz clock, and 192 GB memory running RHEL 7.6 OS. Note that the host plays minimal role in our algorithms, but matters for the baseline.

\begin{table}[h]
\caption{Input graphs (Original Static Graphs)}
\centering
\small

\setlength{\tabcolsep}{4pt}
\begin{tabular}{l|r|r|r|r|r}
\hline
\textbf{Dataset}  & \multicolumn{1}{c|}{\textbf{$|$V$|$}} & \multicolumn{1}{c|}{\textbf{$|$E$|$}} & \textbf{Source} & \textbf{Sink} & \textbf{Flow} \\
               & \multicolumn{1}{c|}{($\times 10^6$)} & \multicolumn{1}{c|}{($\times 10^6$)} & & & \\
              \hline
Flickr-Links    & 1.7  & 15.6  & 397    & 1319705  & 521236  \\ 
Pokecwt         & 1.6  & 30.6  & 5866   & 5934     & 437500  \\ 
Stack Overflow  & 2.6  & 36.2  & 17034  & 22656    & 1117553 \\ 
HollyWood       & 1.1  & 56.3  & 791344 & 599865   & 222925  \\ 
LivJournalwt    & 4.9  & 69.0  & 10009  & 10029    & 680533  \\ 
Wikiwt          & 3.4  & 93.4  & 18646  & 14873    & 250366  \\ 
eu-2015-tpd     & 6.7  & 170.1 & 5521095& 6238766  & 3639651 \\ 
Orkutudwt       & 3.1  & 234.4 & 43607  & 43030    & 1498652 \\ 
UK-2002         & 18.5 & 523.6 & 8504955& 17159800 & 3681010 \\ \hline
\end{tabular}
\label{graph-inputs}
\end{table}

Table~\ref{graph-inputs} lists the input graphs in our experimental setup.
We initialize the edge weights to a random value from 1 to 100. 
We use the graph's average degree ($|E|/|V|$) as a heuristic for \textit{KERNELCYCLES} in Algorithm~\ref{pr_kernel} and ~\ref{pull_kernel}. Global relabeling can perform the relabeling of many vertices at the same time. Hence, if several vertices have finished their pushes in the push relabel kernel, we can save up the relabels for the global relabel.

 We generated the batch update by randomly choosing the existing edges and assigning a higher or lower weight as per the requirement. In order to observe considerable changes in flow, we chose the edges out of the source and into the sink with a higher probability than the other edges. Table~\ref{graph-inputs} also shows the used source and sink ; we chose them to ensure that there is enough processing done by the algorithm. 

 We compare our dynamic implementations with the alternating push-pull implementation mentioned in \cite{Khatri2022} as \textsf{alt-pp}. The dynamic implementation is also compared with the time consumed by the static algorithm after constructing the new graph from the original graph and the batch of edge updates. In all the plots in this section, the x-axis represents the update percentage (1\% to 10\%), while the y-axis is execution time in seconds. 
It should be noted that 1\% itself involves a large change to the underlying graph. For instance, this means, 0.156 million changes for the smallest graph \textsf{Flickr-Links} and 5.236 million changes to the largest graph \textsf{UK-2002}. 
\\Note: We do not have data for alt-pp for Wikiwt dataset for updates more than $3\%$ for Incremental and Mixed as the code timed out during experimentation.

\subsection{Incremental Updates}
\begin{figure*}[t]
    \centering
    \begin{subfigure}{0.33\textwidth}
        \includegraphics[width=\linewidth]{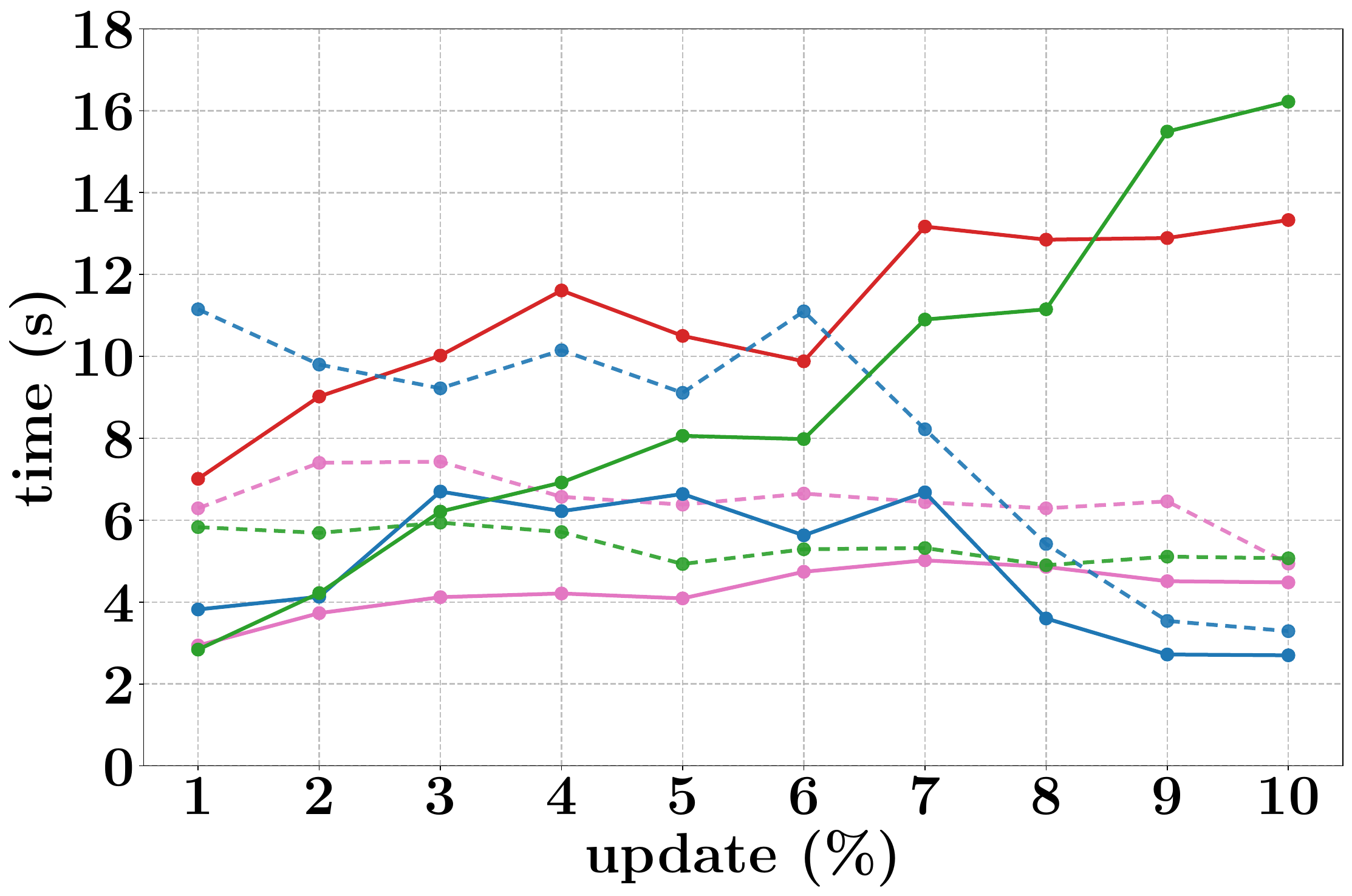}
        \caption{Flickr-Links}
    \end{subfigure}
    \begin{subfigure}{0.33\textwidth}
        \includegraphics[width=\linewidth]{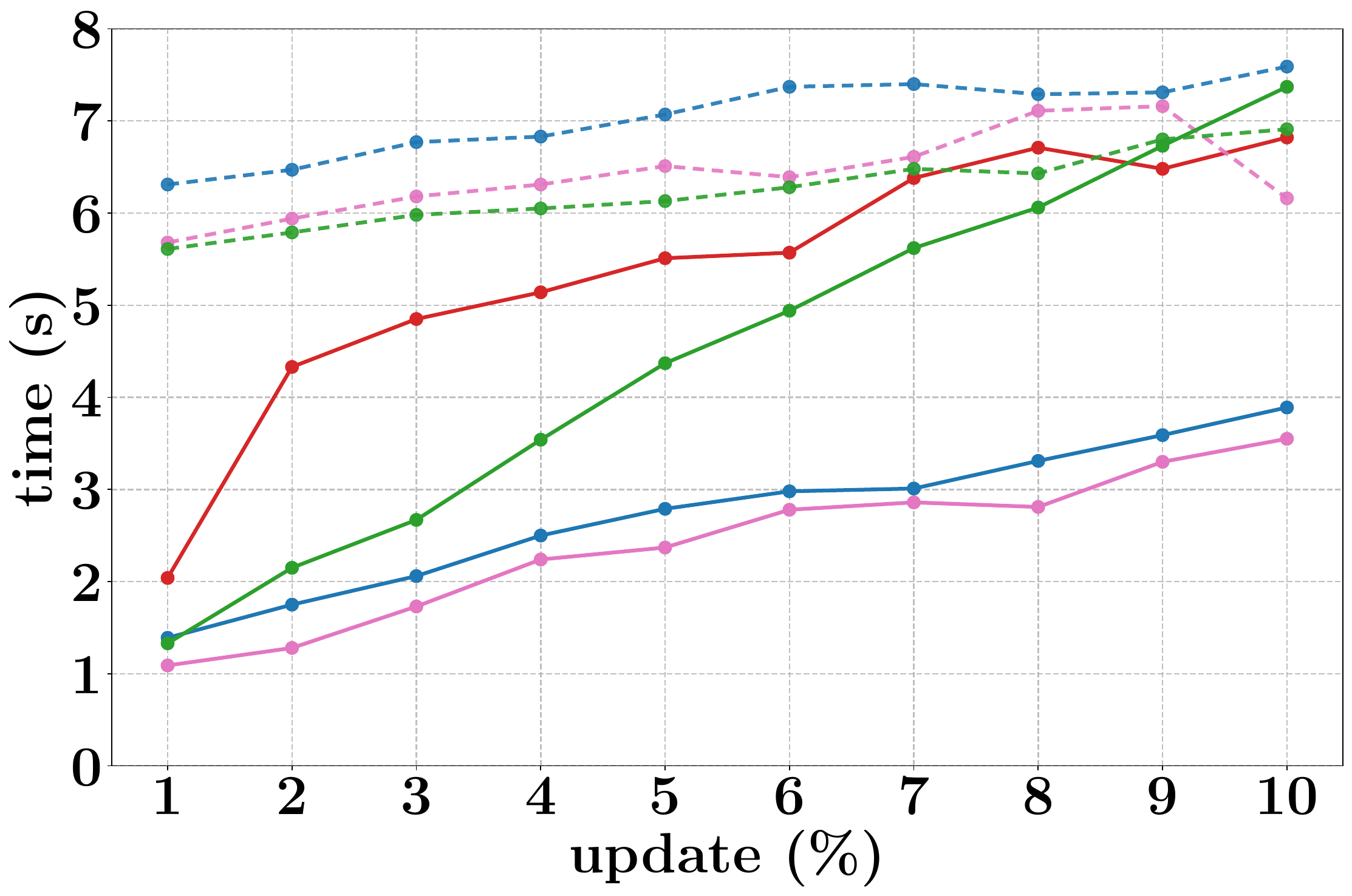}
        \caption{Pokecwt}
    \end{subfigure}
    \begin{subfigure}{0.33\textwidth}
        \includegraphics[width=\linewidth]{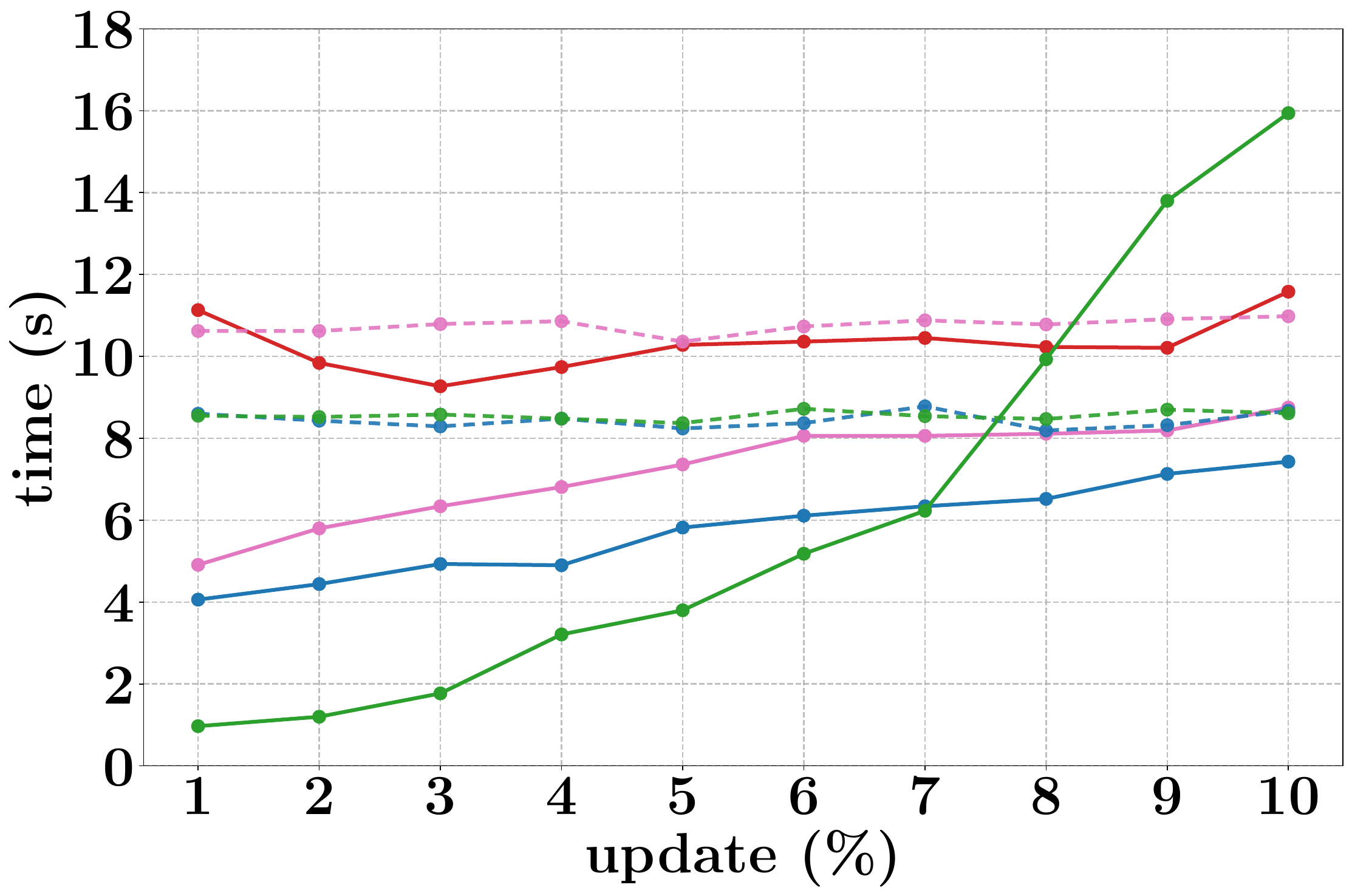}
        \caption{Stack-overflow }
    \end{subfigure}

    \begin{subfigure}{0.33\textwidth}
        \includegraphics[width=\linewidth]{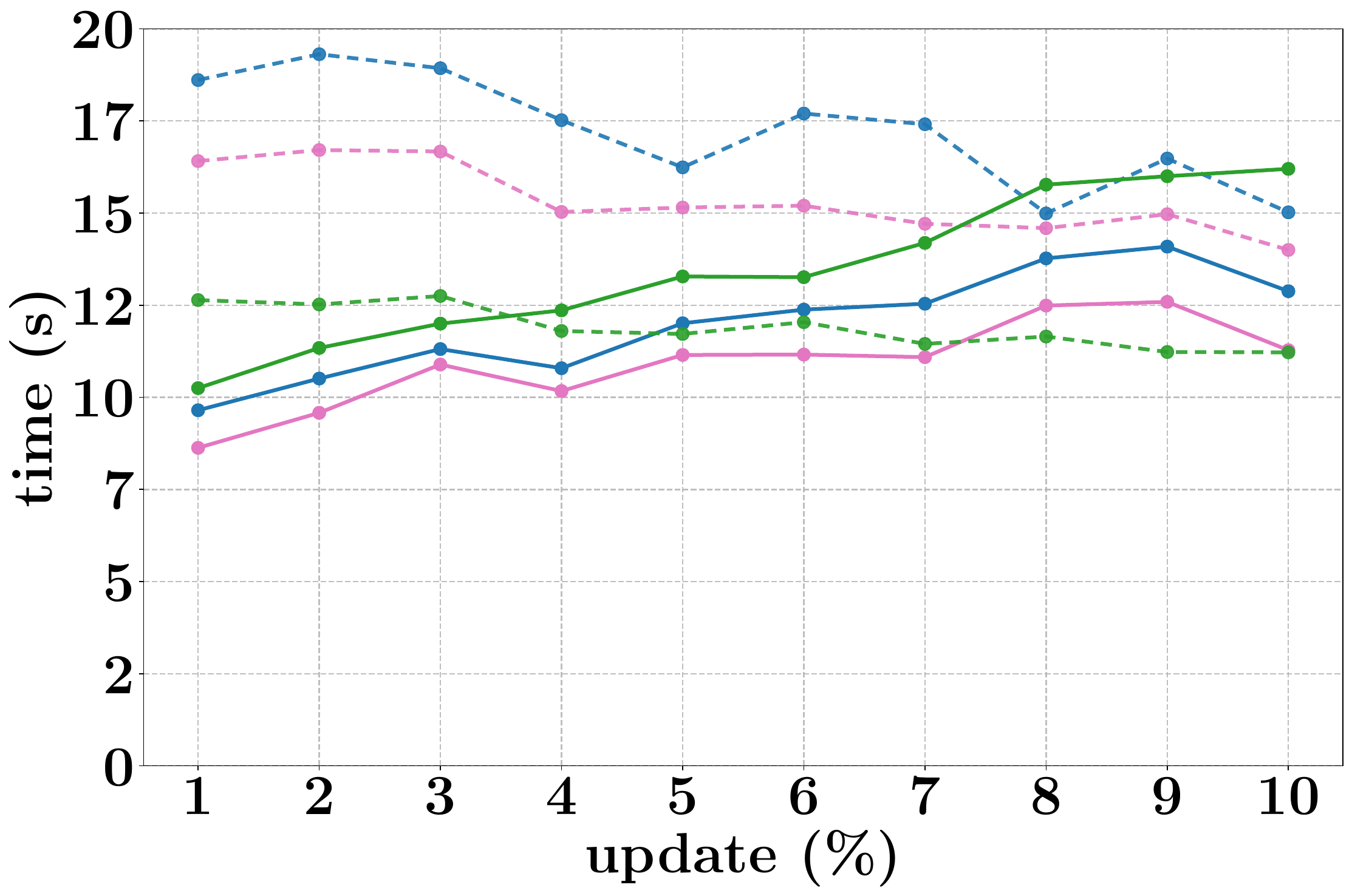}
        \caption{HollyWood }
    \end{subfigure}
    \begin{subfigure}{0.33\textwidth}
        \includegraphics[width=\linewidth]{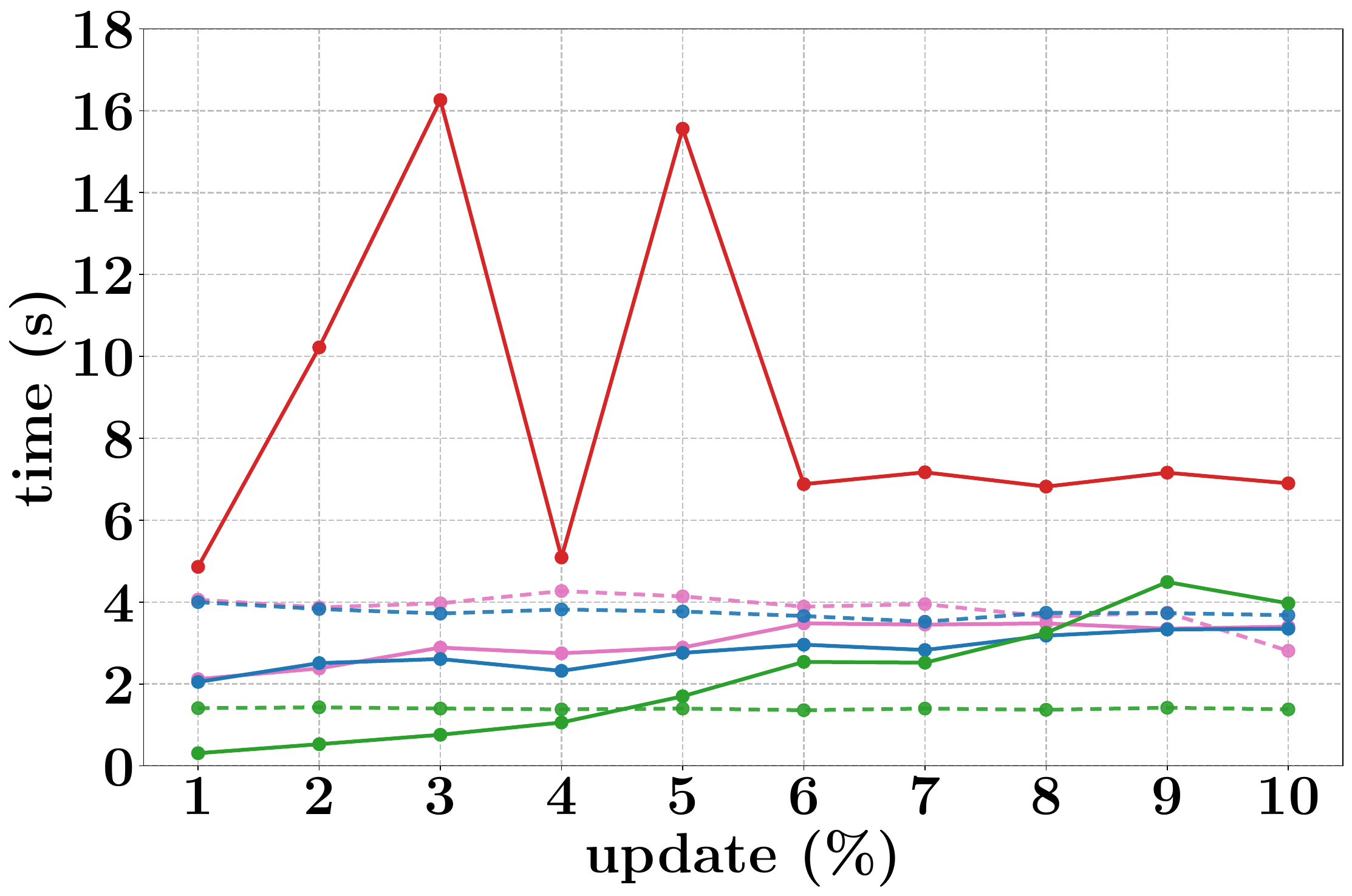}
        \caption{LivJournalwt}
    \end{subfigure}
    \begin{subfigure}{0.33\textwidth}
        \includegraphics[width=\linewidth]{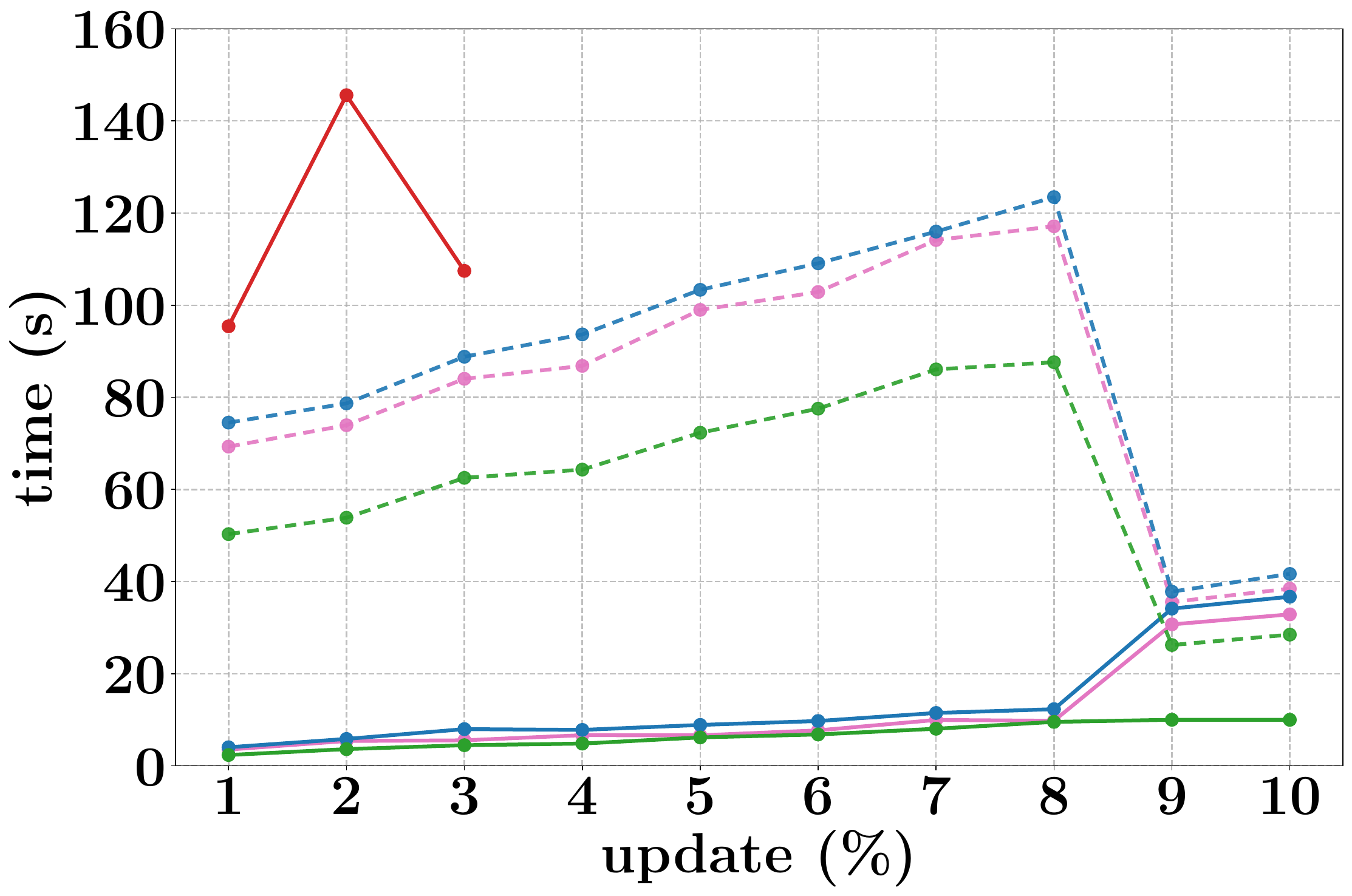}
        \caption{Wikiwt}
    \end{subfigure}

    \begin{subfigure}{0.33\textwidth}
        \includegraphics[width=\linewidth]{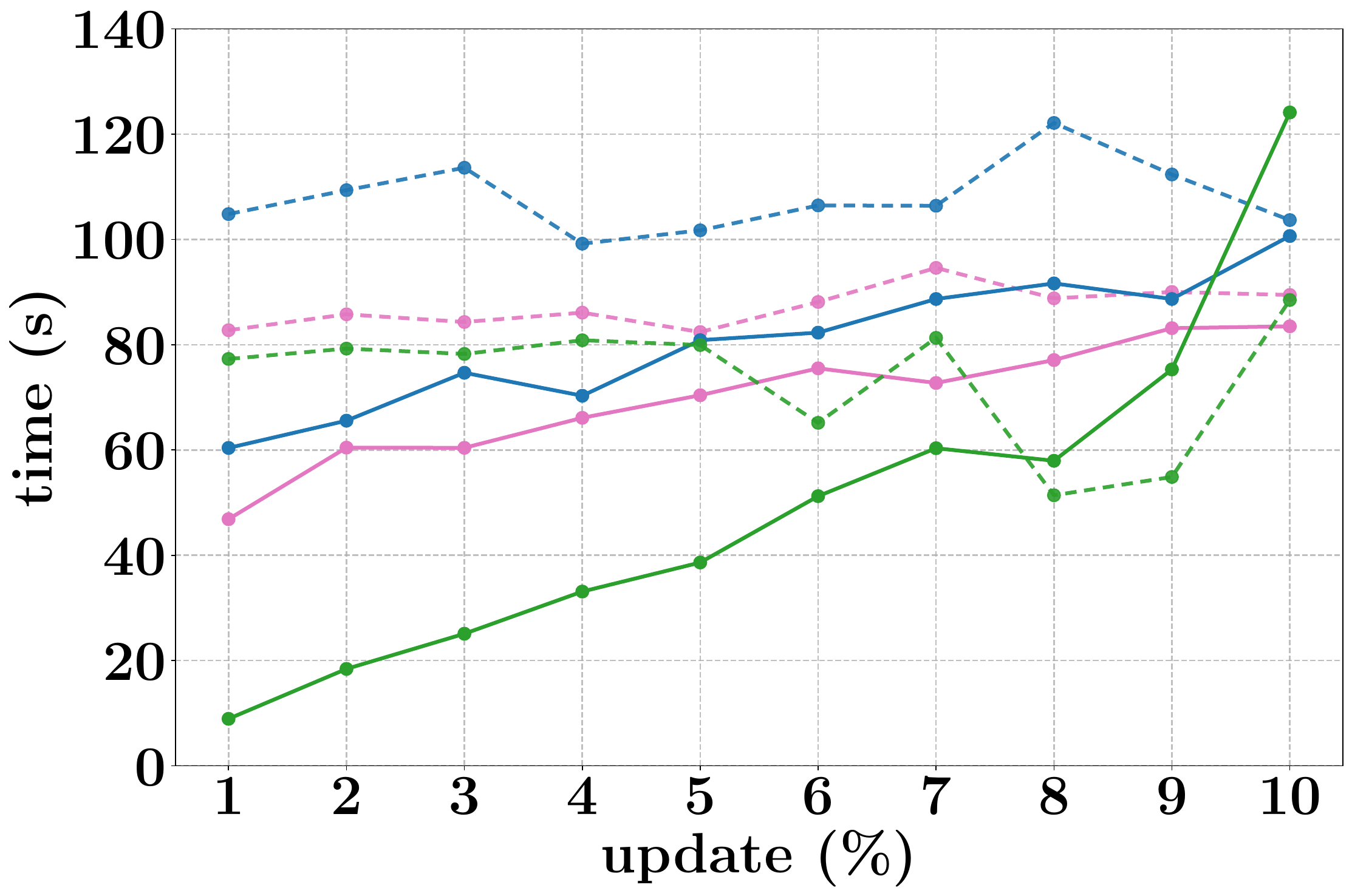}
        \caption{eu-2015-tpd }
    \end{subfigure}
    \begin{subfigure}{0.33\textwidth}
        \includegraphics[width=\linewidth]{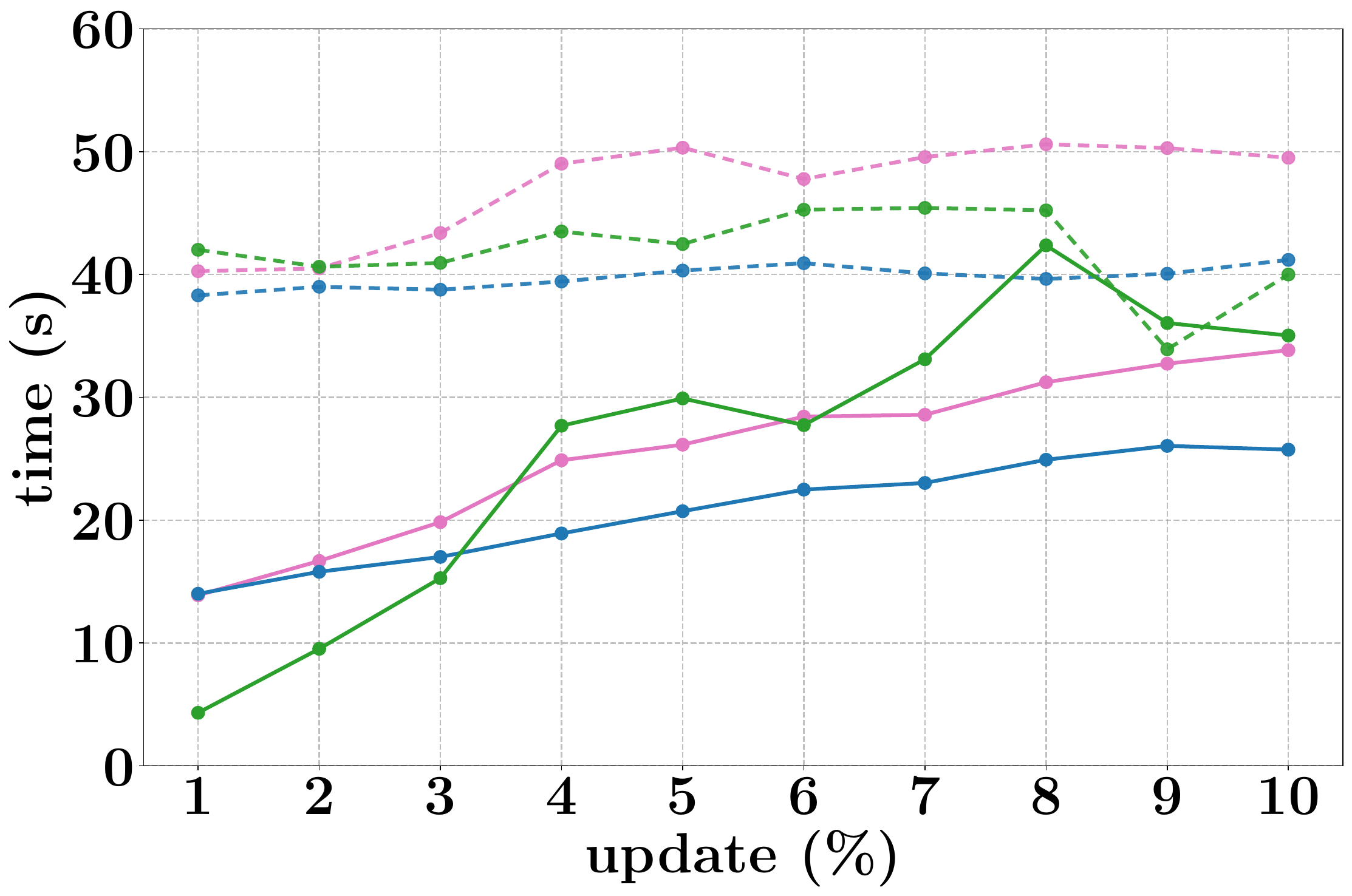}
        \caption{Orkutudwt}
    \end{subfigure}
    \begin{subfigure}{0.33\textwidth}
        \includegraphics[width=\linewidth]{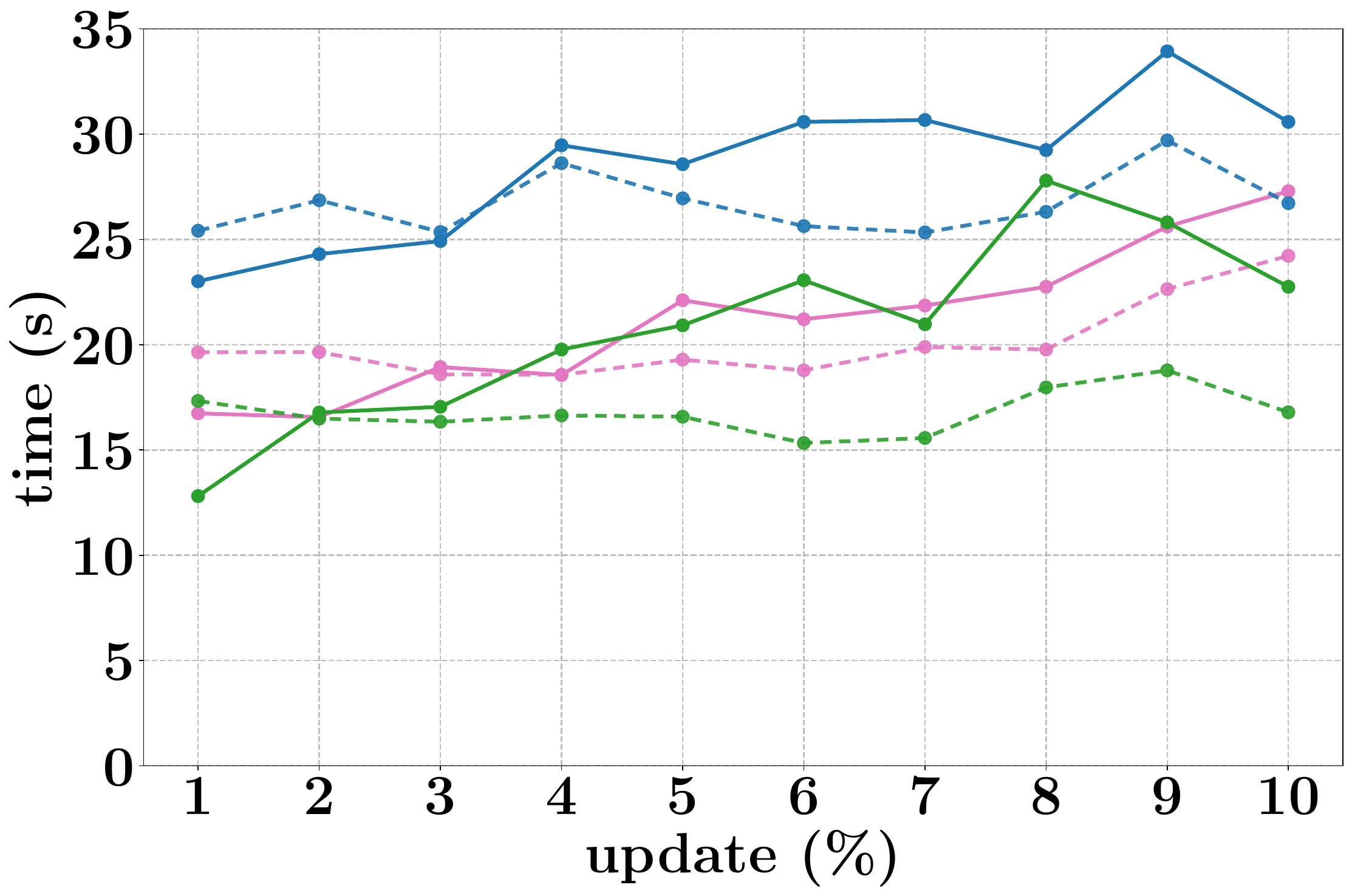}
        \caption{UK-2002}
    \end{subfigure}

\caption{
Performance on Incremental updates. Legend:
\protect\tikz[baseline=-0.6ex]\protect\node[fill=tabred,draw,circle,inner sep=2pt]{};
~alt-pp,\ 
\protect\tikz[baseline=-0.6ex]\protect\node[fill=tabgreen,draw,circle,inner sep=2pt]{};
~Push--Pull,\ 
\protect\tikz[baseline=-0.6ex]\protect\node[fill=tabblue,draw,circle,inner sep=2pt]{};
~Topology-driven,\ 
\protect\tikz[baseline=-0.6ex]\protect\node[fill=tabpink,draw,circle,inner sep=2pt]{};
~Data-driven.
\\Solid lines = dynamic, dashed lines = static.
}
    \label{fig:set2}
\end{figure*}
The results for incremental updates are presented in Figure~\ref{fig:set2}. Here, the new capacities of all the batch edges are higher than the original capacity. Such updates reflect the capability of sending more flow from the source or overflowing vertices to the sink. 

On medium to large graphs such as \textsf{Stack Overflow}, \textsf{eu-2015-tpd}, and \textsf{Orkutudwt}, PP exhibits the highest improvements at smaller update ratios. It achieves $11.5\times$, $8.2\times$, and $5.2\times$ speedups over \textsf{alt-pp} for $1$–$3\%$ updates on \textsf{Stack Overflow}, and $8.8\times$, $7\times$, and $4.7\times$ gains over static. On \textsf{eu-2015-tpd}, PP achieves up to $8.6\times$ at $1\%$ and maintains an average $3.3\times$ improvement till $7\%$ updates, while on \textsf{Orkutudwt}, it records $8.9\times$, $4.1\times$, and $2.5\times$ improvements for $1$–$3\%$ updates against static. 

The \textsf{Data-Driven } variant provides consistent advantages across smaller social graphs. Against static recomputation, it achieves average improvements of $2.9\times$ on \textsf{Pokecwt} till $10\%$ (peaking at $5.2\times$ at $1\%$), $1.5\times$ on \textsf{Flickr-Links} till $7\%$, and $1.2\times$ on \textsf{Hollywood} till $7\%$. Against \textsf{alt-pp}, data driven remains dominant on all small and medium graphs, with average gains of $2.6\times$ on \textsf{Flickr-Links}, $2.3\times$ on \textsf{Pokecwt}, and $139\times$ on \textsf{Hollywood} (alt-pp thus not plotted). On \textsf{LivJournalwt}, PP provides an $8.1\times$ average speedup over \textsf{alt-pp} and $4.6\times$ at $1\%$ over static recomputation. On \textsf{Wikiwt}, all variants perform exceptionally well, with PP achieving $21.3\times$, $14.8\times$, and $13.9\times$ improvements over static at $1$–$3\%$ updates, and maintaining an $11.2\times$ average till $10\%$. For the largest dataset, \textsf{UK-2002}, static recomputation dominates overall, with PP offering a modest $1.4\times$ improvement at $1\%$. 

Across all datasets, PP excels at smaller update ratios on high-volume graphs while the Data-Driven variant maintains steady gains across smaller networks.

\subsection{Decremental Updates}
\begin{figure*}[t]
    \centering
    \begin{subfigure}{0.33\textwidth}
        \includegraphics[width=\linewidth]{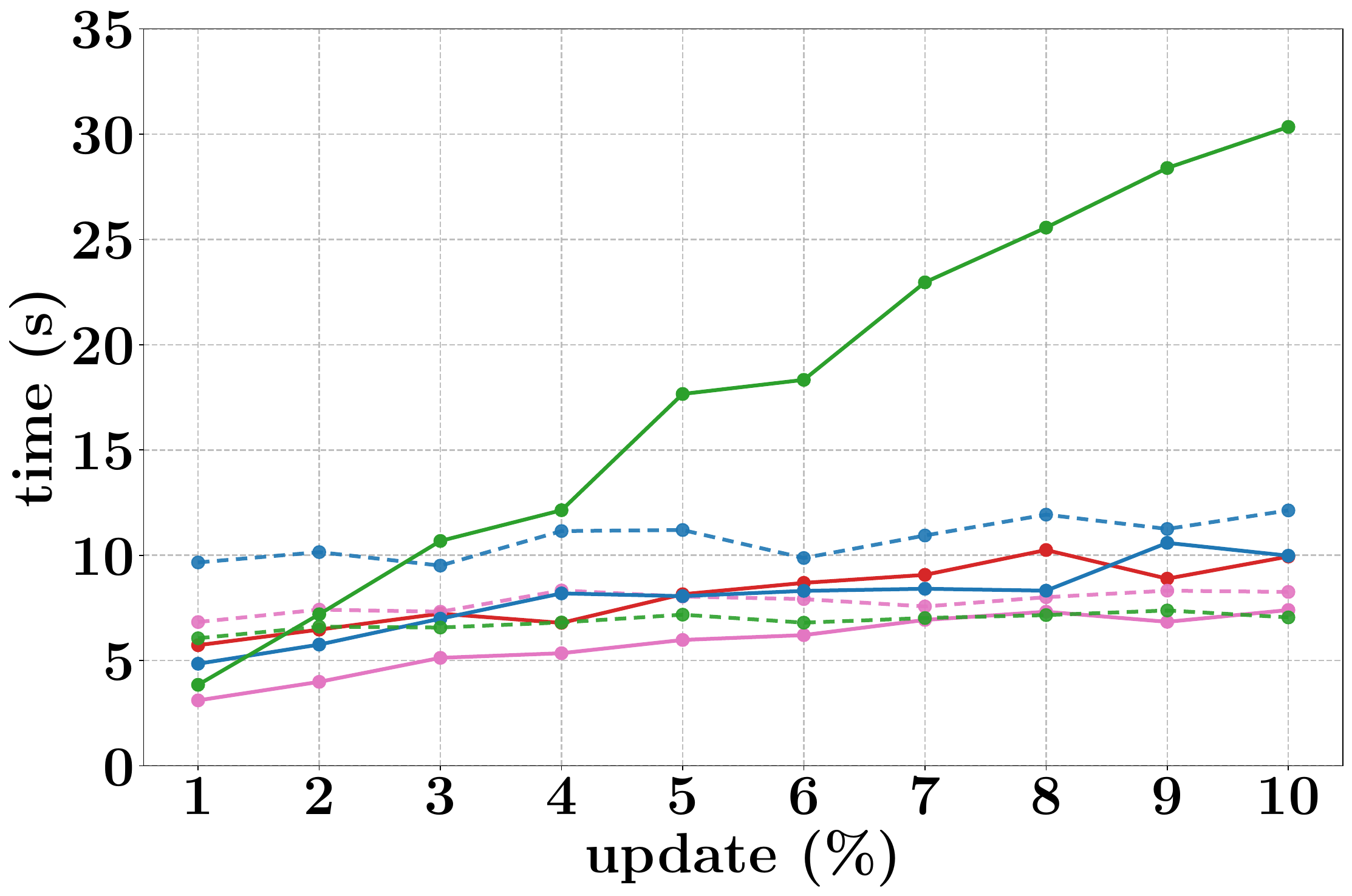}
        \caption{Flickr-Links}
    \end{subfigure}
    \begin{subfigure}{0.33\textwidth}
        \includegraphics[width=\linewidth]{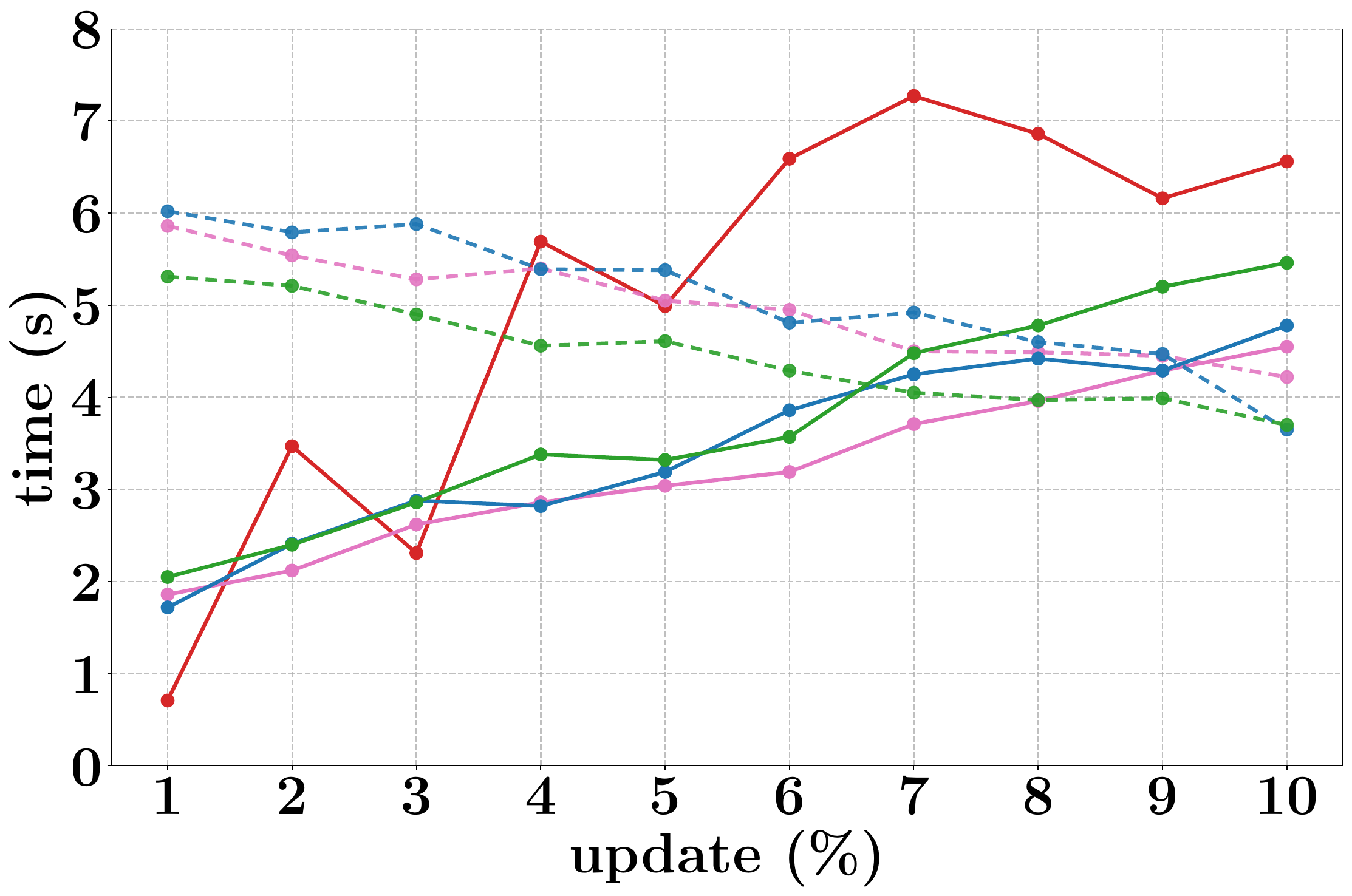}
        \caption{Pokecwt}
    \end{subfigure}
    \begin{subfigure}{0.33\textwidth}
        \includegraphics[width=\linewidth]{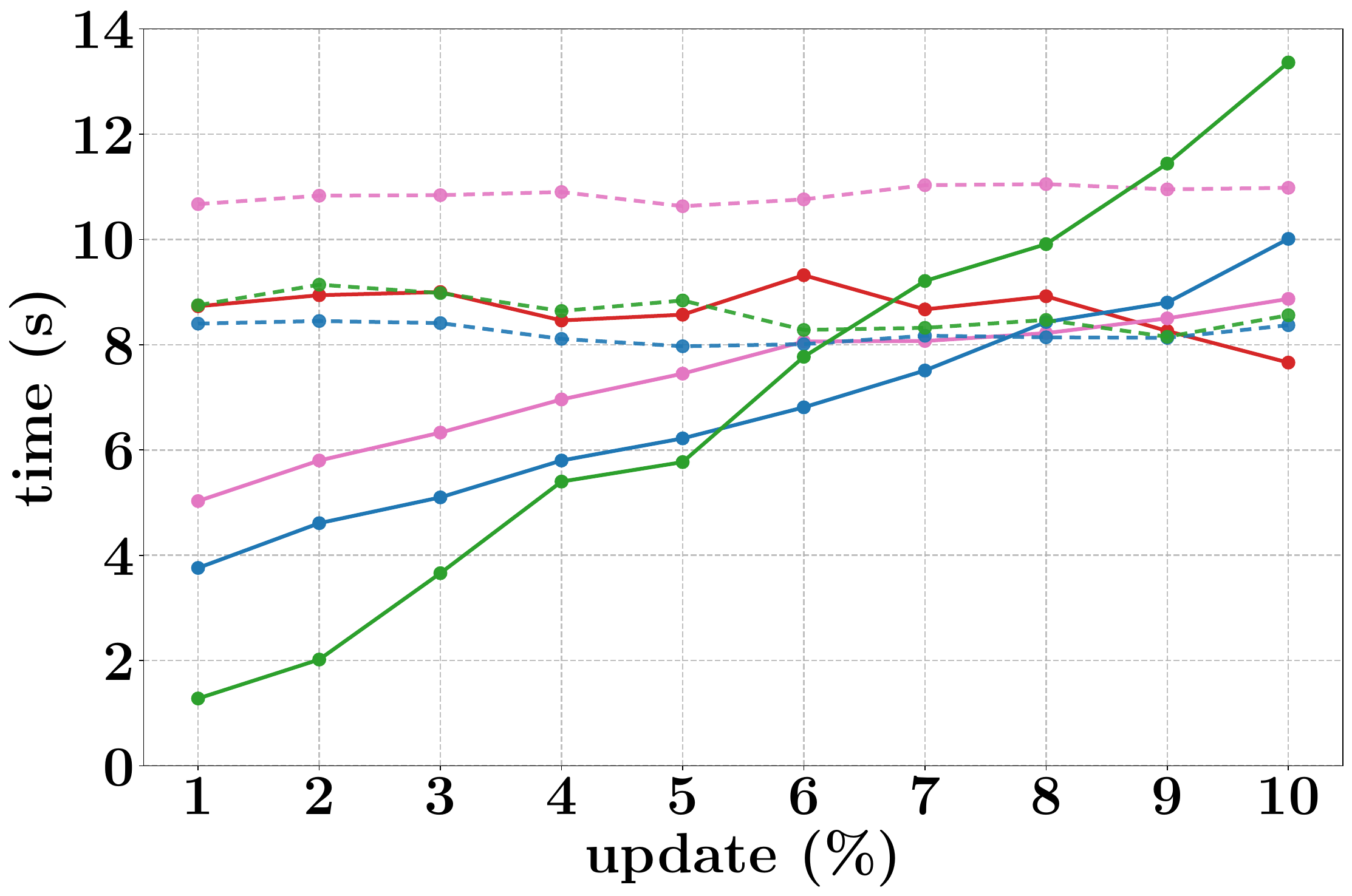}
        \caption{Stack-overflow }
    \end{subfigure}

    \begin{subfigure}{0.33\textwidth}
        \includegraphics[width=\linewidth]{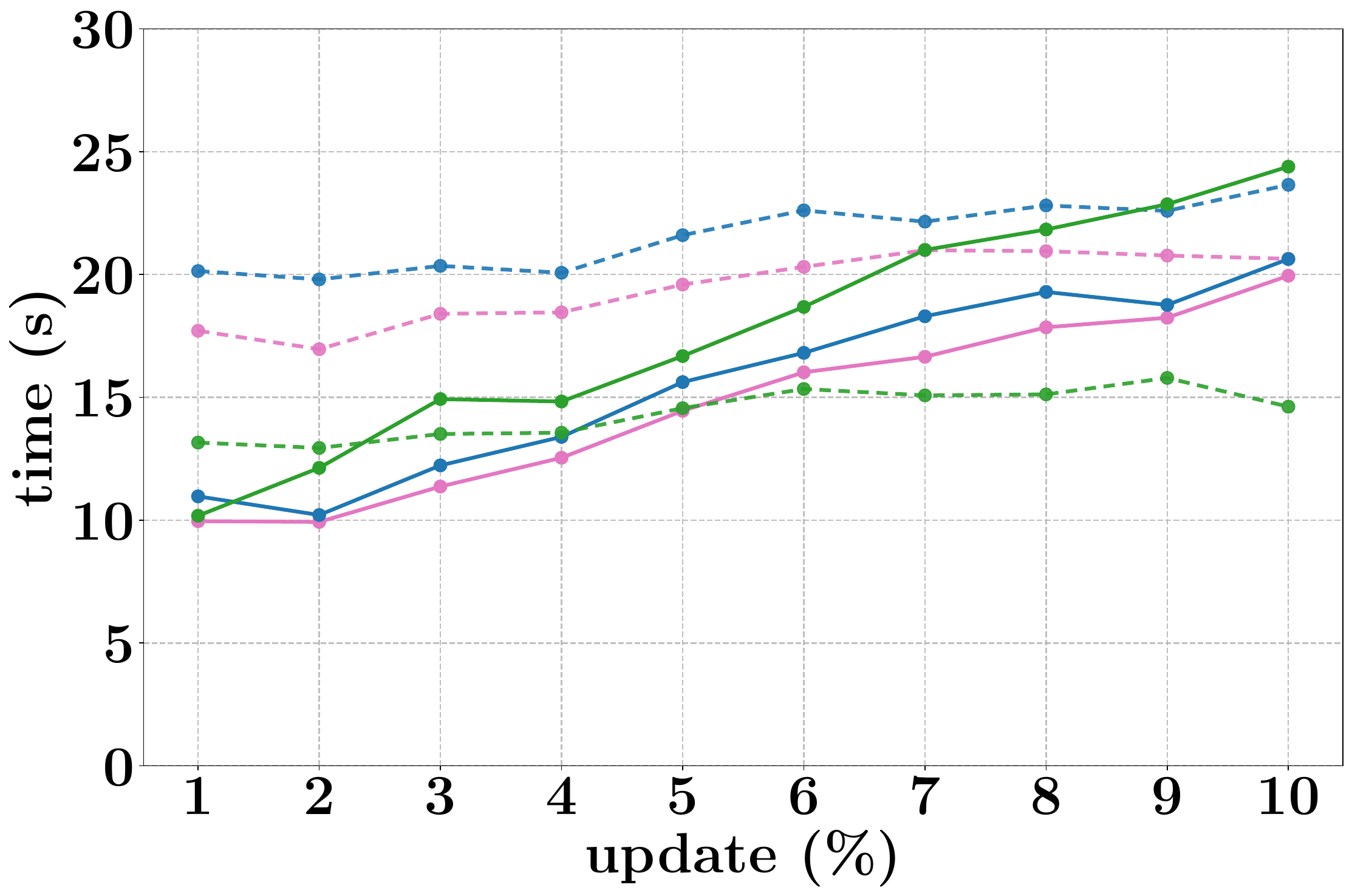}
        \caption{HollyWood }
    \end{subfigure}
    \begin{subfigure}{0.33\textwidth}
        \includegraphics[width=\linewidth]{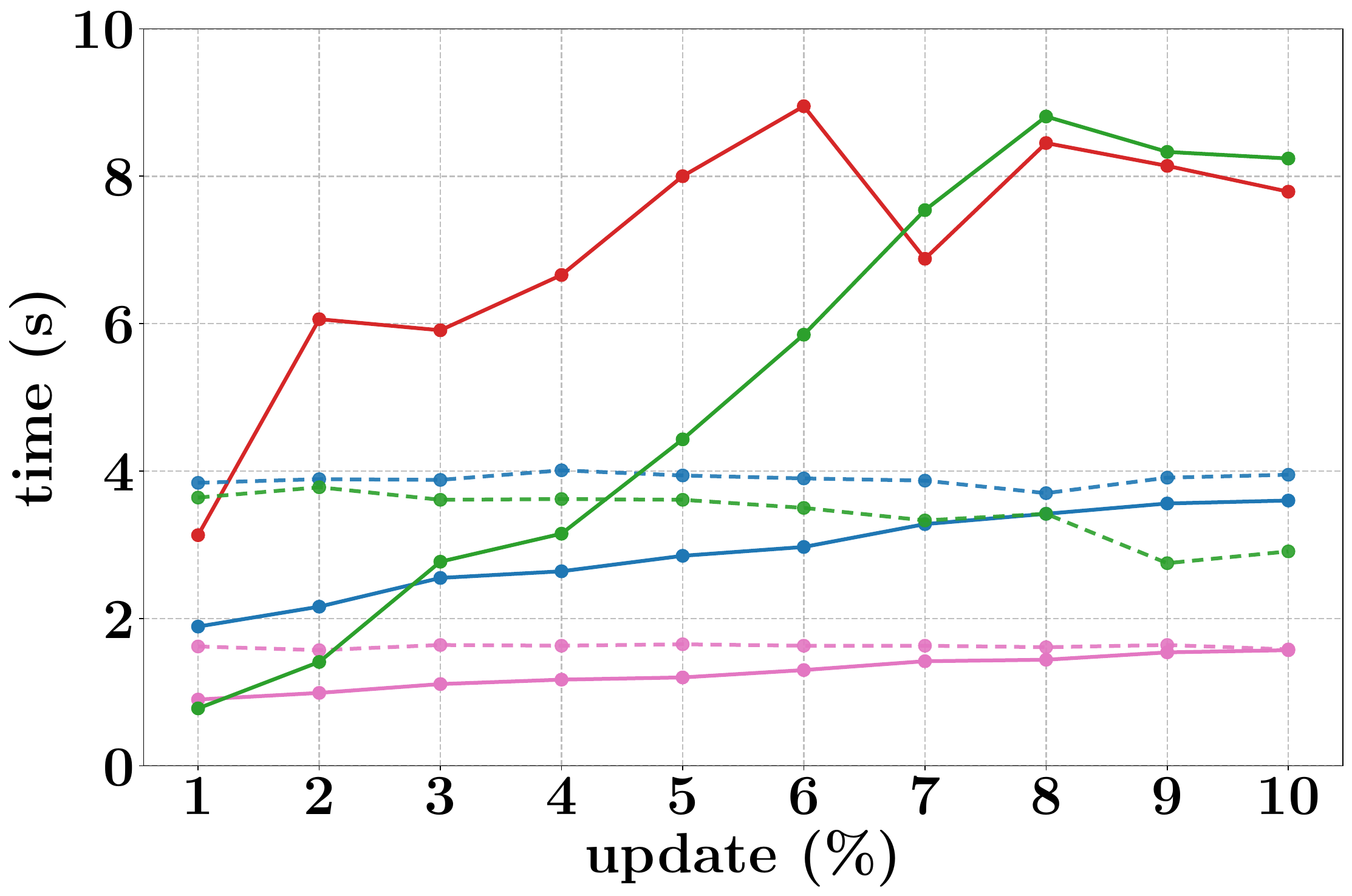}
        \caption{LivJournalwt}
    \end{subfigure}
    \begin{subfigure}{0.33\textwidth}
        \includegraphics[width=\linewidth]{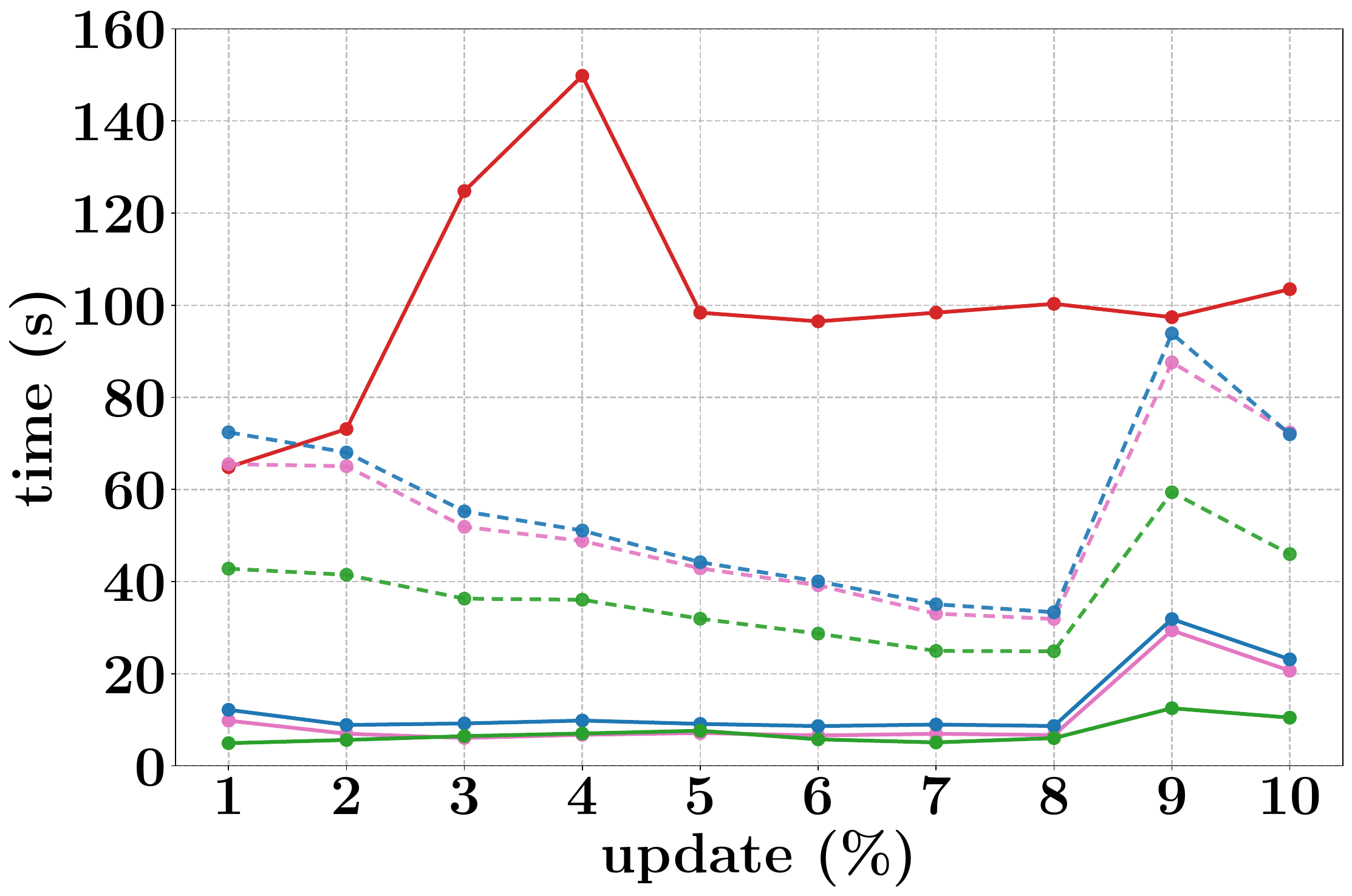}
        \caption{Wikiwt}
    \end{subfigure}

    \begin{subfigure}{0.33\textwidth}
        \includegraphics[width=\linewidth]{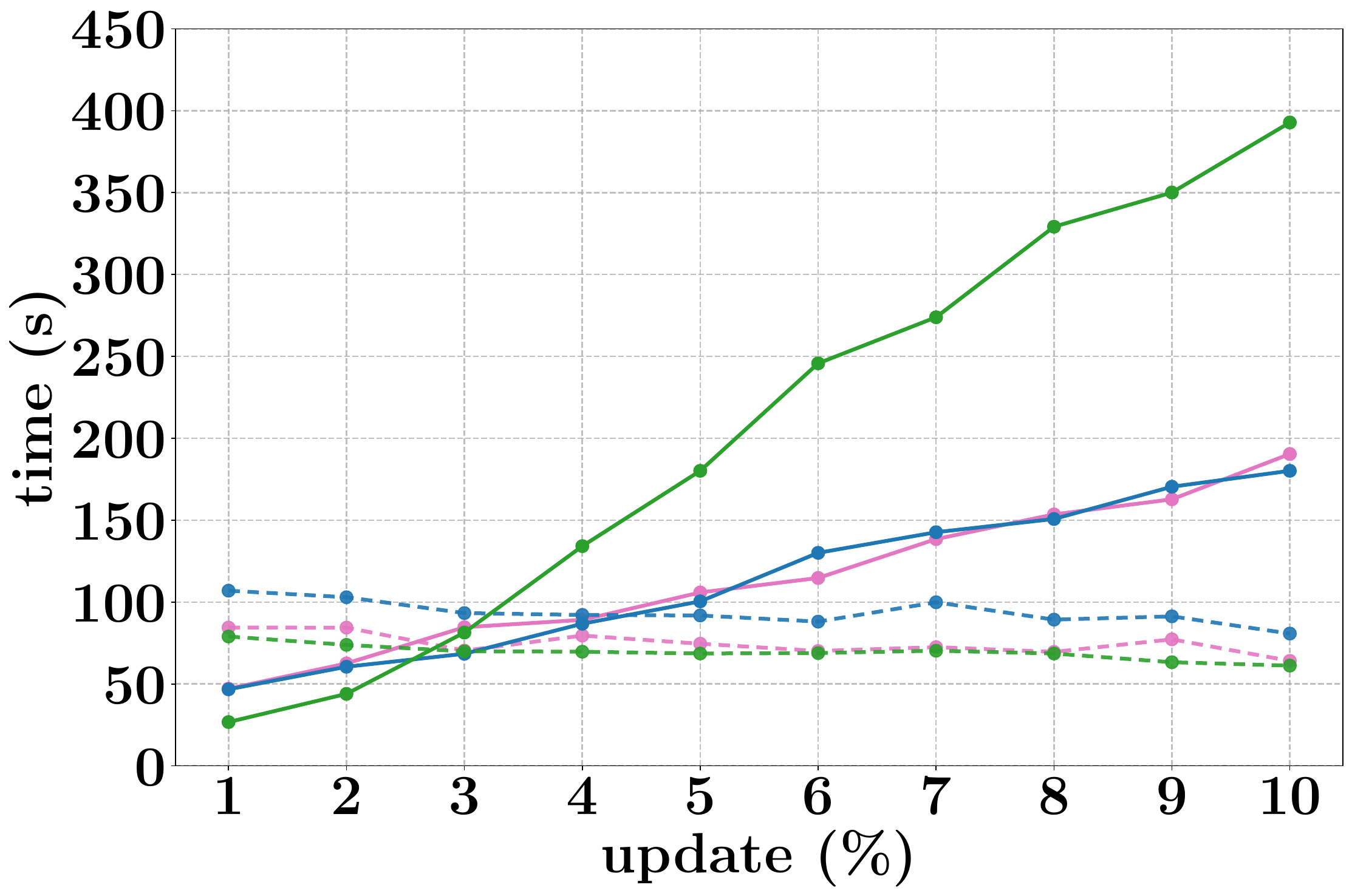}
        \caption{eu-2015-tpd }
    \end{subfigure}
    \begin{subfigure}{0.33\textwidth}
        \includegraphics[width=\linewidth]{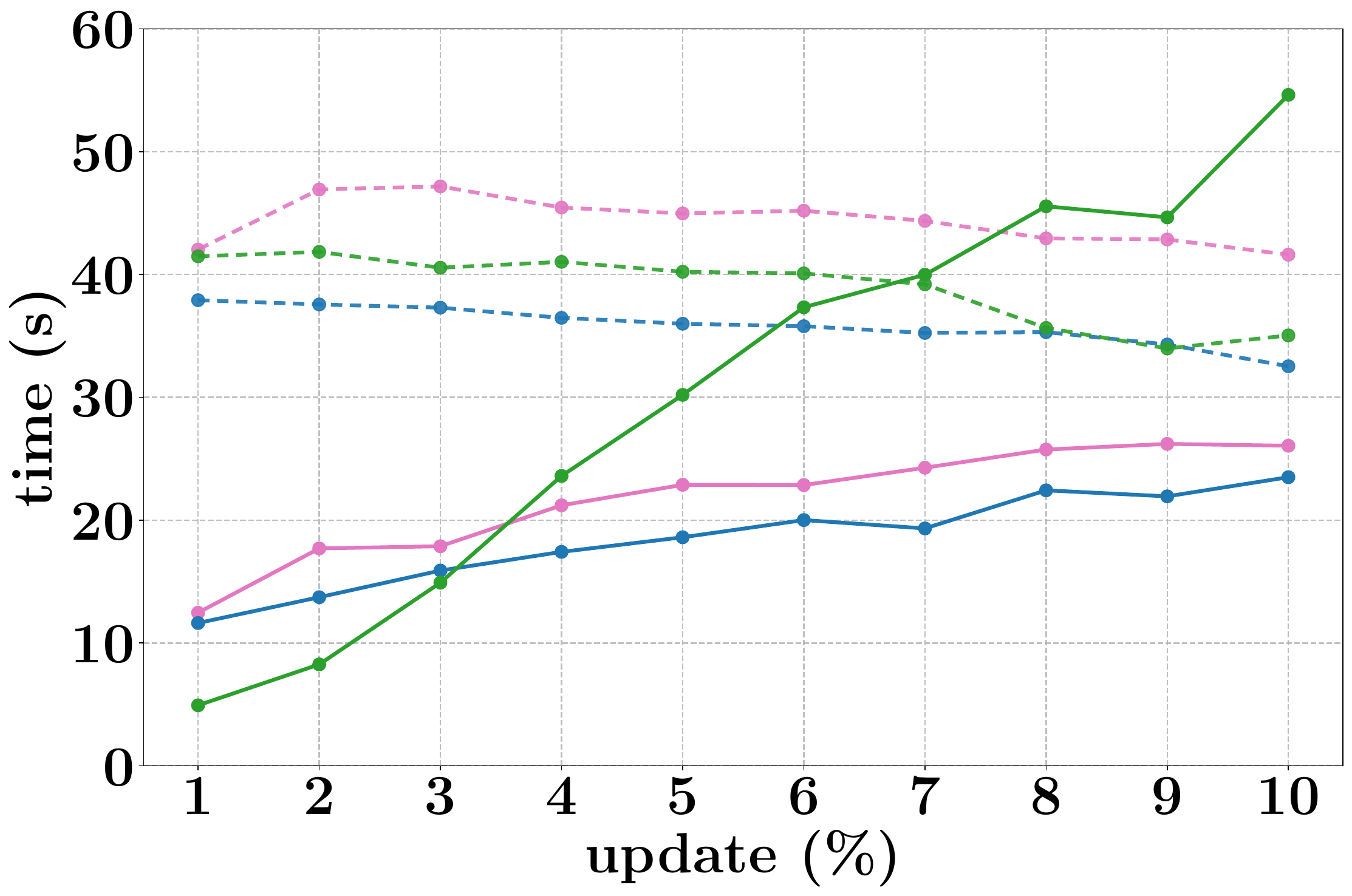}
        \caption{Orkutudwt}
    \end{subfigure}
    \begin{subfigure}{0.33\textwidth}
        \includegraphics[width=\linewidth]{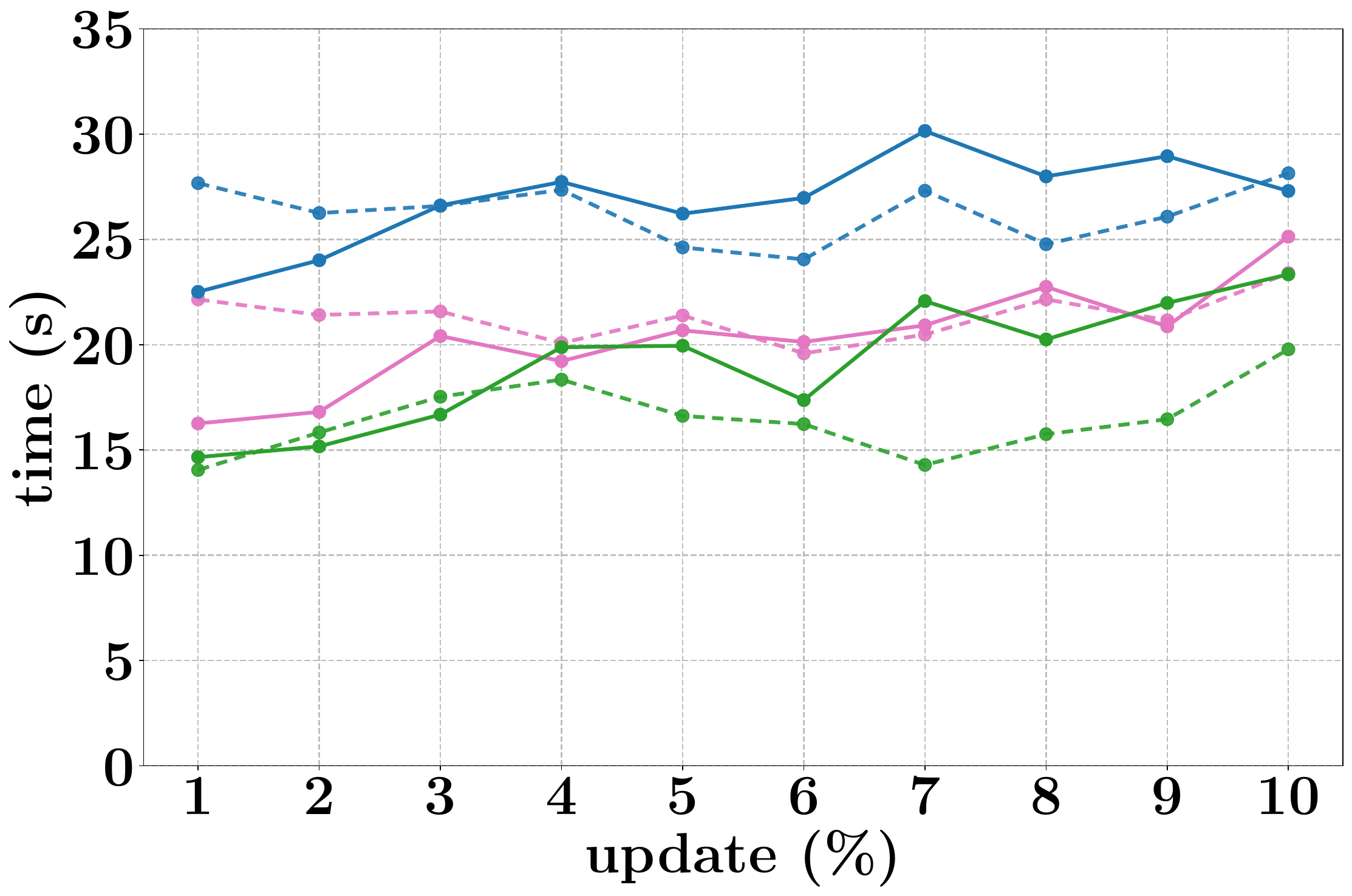}
        \caption{UK-2002}
    \end{subfigure}

\caption{
Performance on Decremental updates. Legend:
\protect\tikz[baseline=-0.6ex]\protect\node[fill=tabred,draw,circle,inner sep=2pt]{};
~alt-pp,\ 
\protect\tikz[baseline=-0.6ex]\protect\node[fill=tabgreen,draw,circle,inner sep=2pt]{};
~Push--Pull,\ 
\protect\tikz[baseline=-0.6ex]\protect\node[fill=tabblue,draw,circle,inner sep=2pt]{};
~Topology-driven,\ 
\protect\tikz[baseline=-0.6ex]\protect\node[fill=tabpink,draw,circle,inner sep=2pt]{};
~Data-driven.
\\Solid lines = dynamic, dashed lines = static.
}
    \label{fig:set1}
\end{figure*}
The results for decremental updates are presented in Figure~\ref{fig:set1}. In these updates, the new capacities of all the batch edges are less than their original capacity. Such updates reflect the capability of sending back flow to the source or taking back from the sink. 

The \textsf{Push--Pull Streams (PP)} implementation demonstrates the most significant improvements during the initial update ratios on medium to large datasets such as \textsf{Stack Overflow}, \textsf{eu-2015-tpd}, and \textsf{Orkutudwt}. For instance, on \textsf{Stack Overflow}, PP achieves up to $6.8\times$ improvement at $1\%$ updates and maintains a $3.4\times$ average speedup over \textsf{alt-pp} for updates up to $5\%$, while providing comparable gains over the static baseline. Similarly, on \textsf{eu-2015-tpd}, PP offers $3\times$ and $1.7\times$ speedups at $1\%$ and $2\%$ updates, respectively, against static. On \textsf{Orkutudwt}, PP achieves $7.7\times$, $4.5\times$, and $2.5\times$ improvements for $1$–$3\%$ updates against static baselines.

The \textsf{Data-Driven} variant yields the most consistent gains across small and medium-sized graphs, outperforming \textsf{alt-pp} and static recomputation on \textsf{Hollywood}, \textsf{LivJournalwt}, \textsf{Pokecwt}, and \textsf{Flickr-Links}. For example, against static baselines, it achieves an average $1.4\times$ improvement till $9\%$ update on \textsf{LivJournalwt}, $1.4\times$ till $6\%$ update on \textsf{Flickr-Links}, $1.2\times$ till $5\%$ update on \textsf{Hollywood}, and $1.8\times$ till $7\%$ update on \textsf{Pokecwt}. It shows exceptionally high speedups—$194\times$ for data-driven
and $181\times$ for topology driven—against alt-pp, hence alt-pp is omitted from the plot. On \textsf{Wikiwt}, all three variants perform strongly, with PP leading at $15\times$ over \textsf{alt-pp} and an average $5.4\times$ improvement over static. For the largest dataset, \textsf{UK-2002}, static recomputation dominates.

Overall, the \textsf{data-driven} approach offers stable speedups across small and medium graphs, and \textsf{pp} excels at small batch sizes on large graphs—while both consistently outperform \textsf{alt-pp} except in \textsf{Pokecwt} and \textsf{Stack Overflow}, where the latter performs competitively at isolated update ratios.
\subsection{Mixed Updates}
\begin{figure*}[t]
    \centering
    \begin{subfigure}{0.33\textwidth}
        \includegraphics[width=\linewidth]{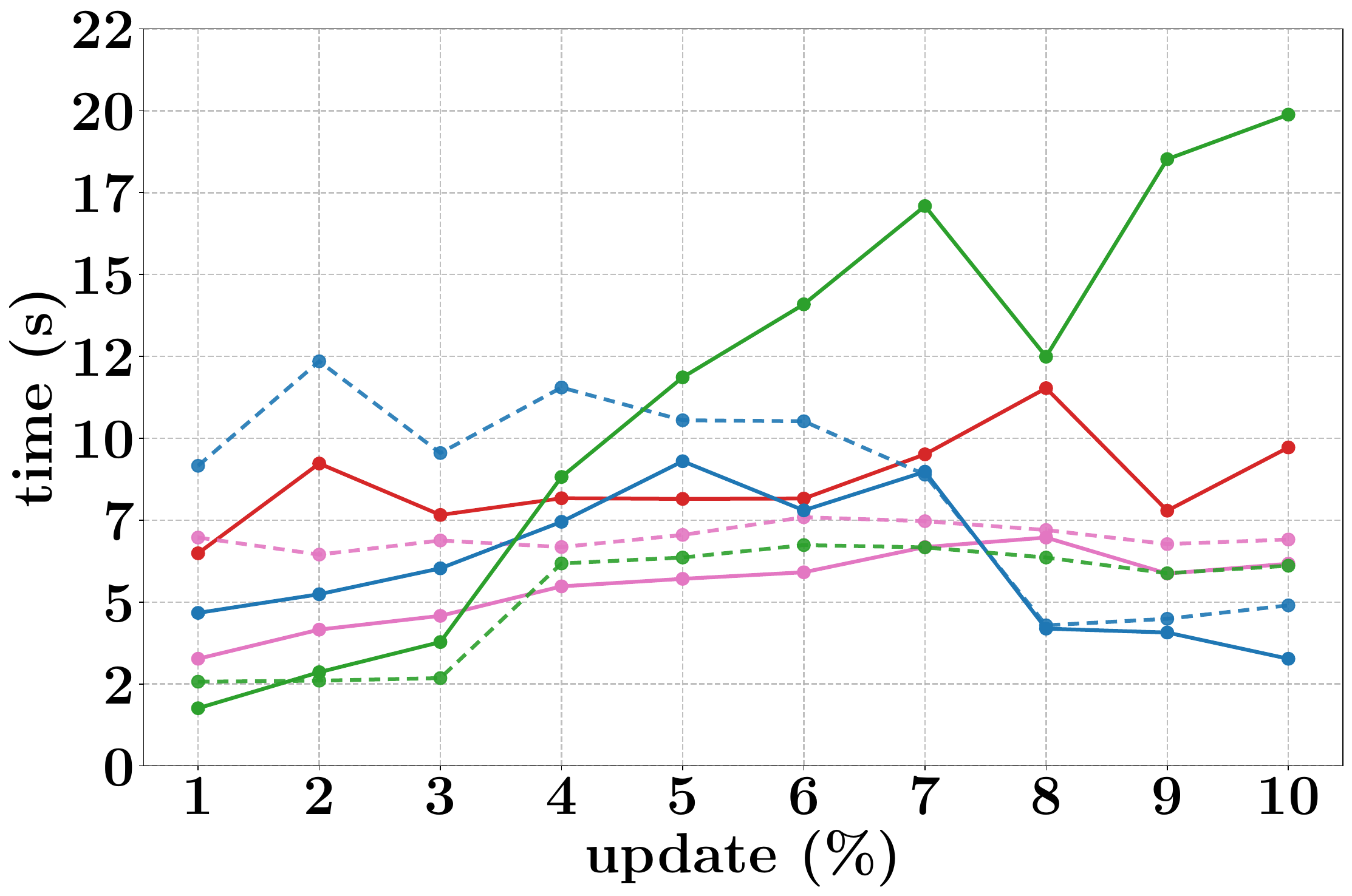}
        \caption{Flickr-Links}
    \end{subfigure}
    \begin{subfigure}{0.33\textwidth}
        \includegraphics[width=\linewidth]{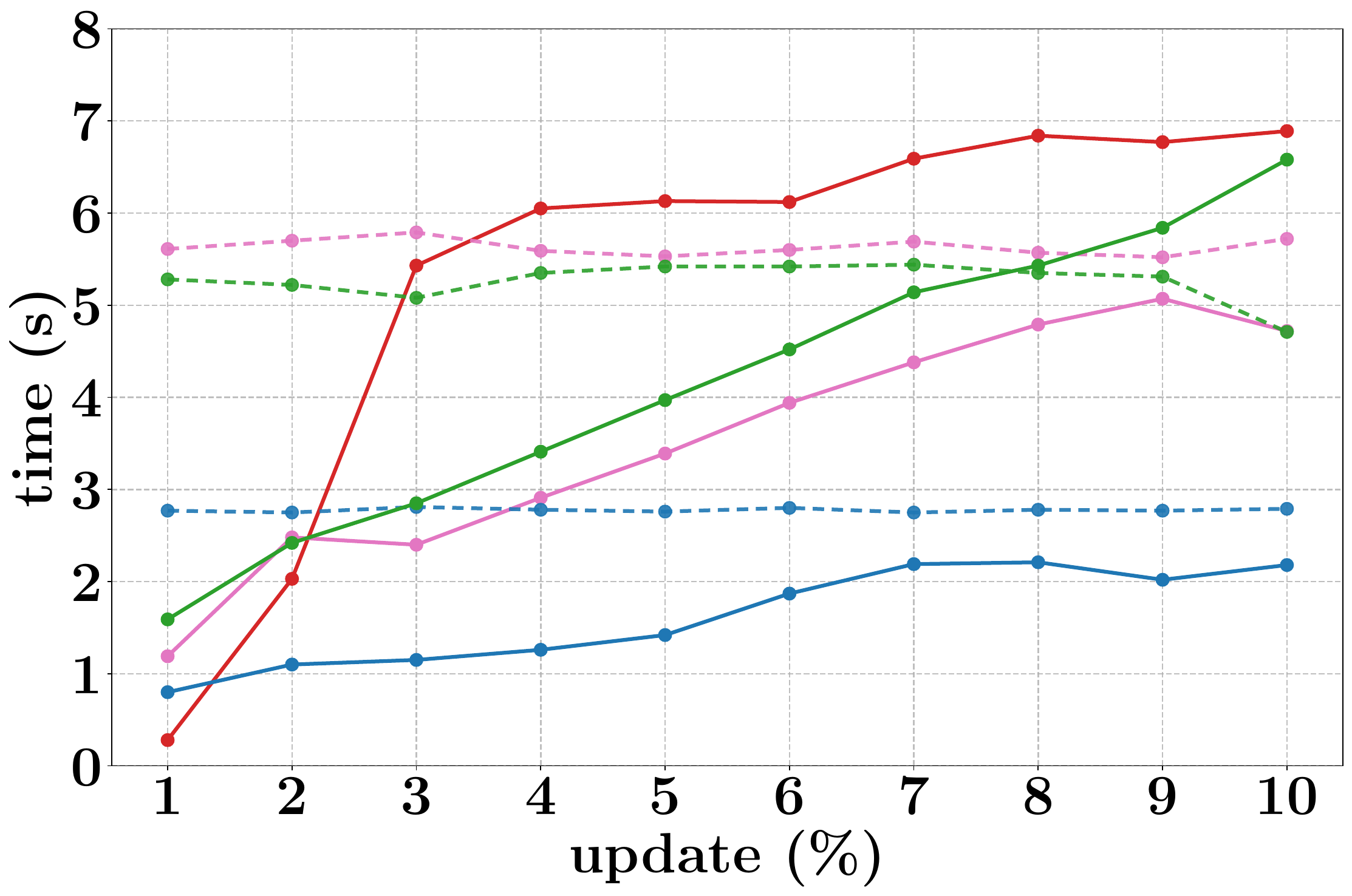}
        \caption{Pokecwt}
    \end{subfigure}
    \begin{subfigure}{0.33\textwidth}
        \includegraphics[width=\linewidth]{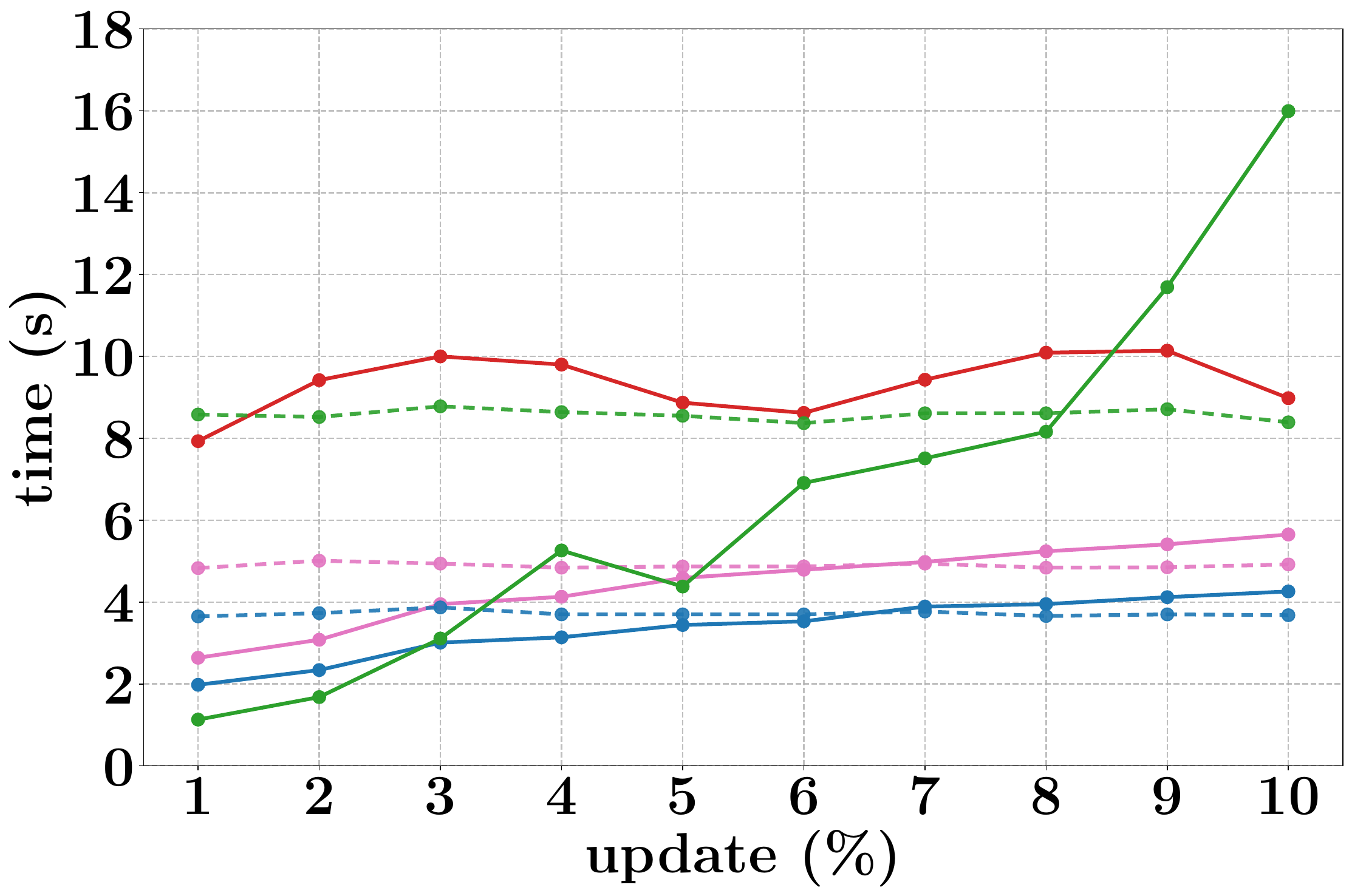}
        \caption{Stack-overflow }
    \end{subfigure}

    \begin{subfigure}{0.33\textwidth}
        \includegraphics[width=\linewidth]{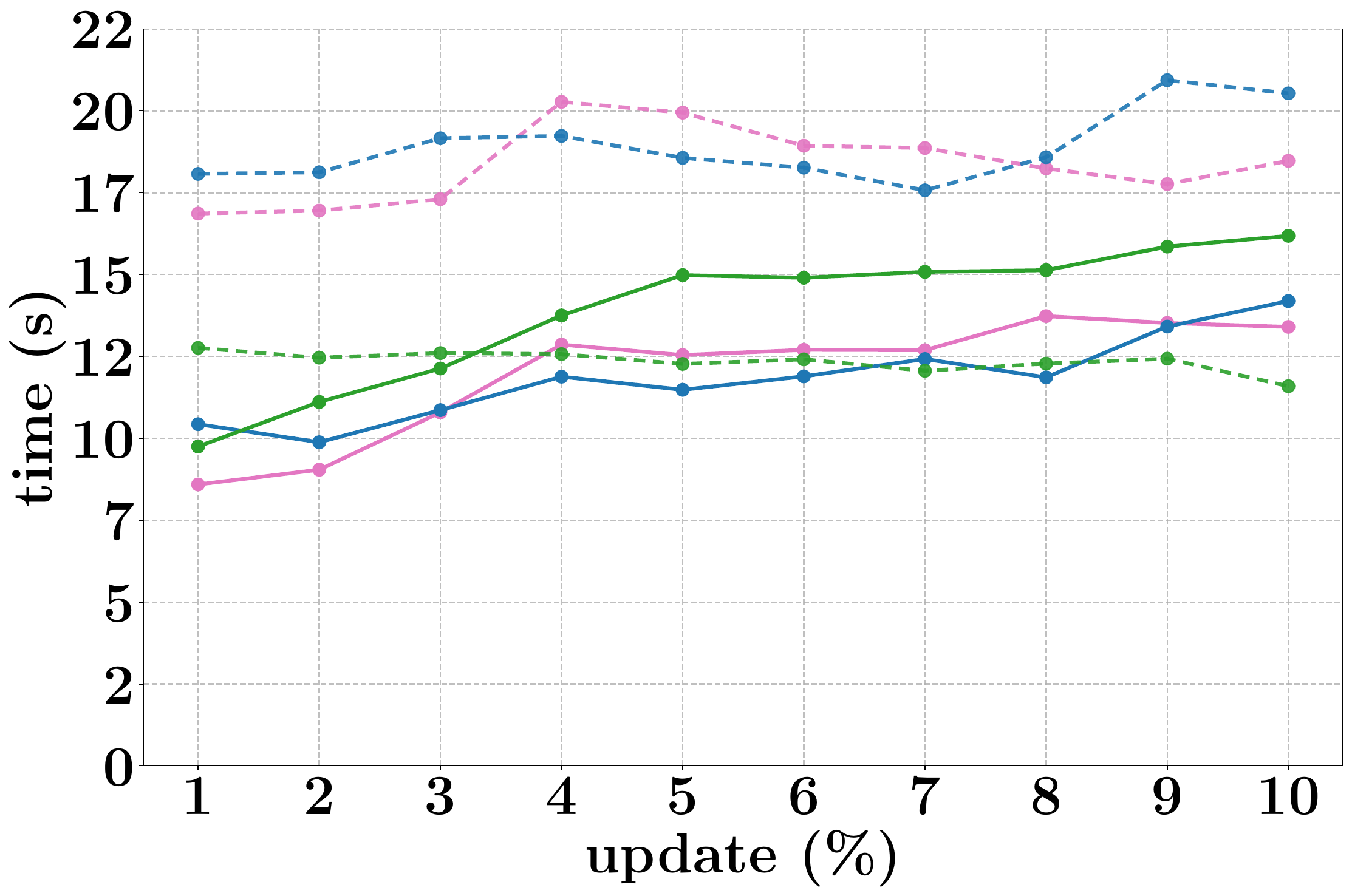}
        \caption{HollyWood }
    \end{subfigure}
    \begin{subfigure}{0.33\textwidth}
        \includegraphics[width=\linewidth]{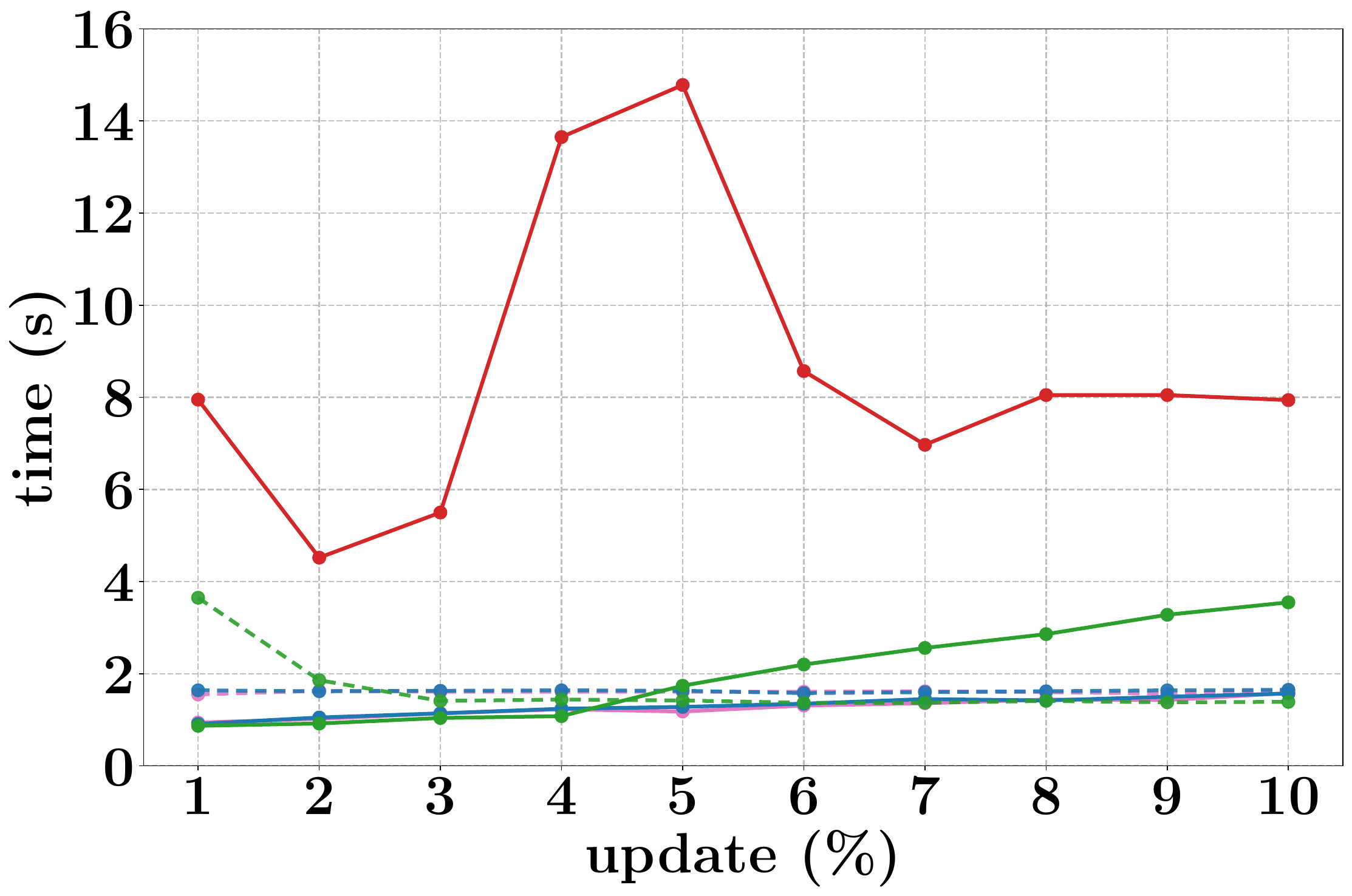}
        \caption{LivJournalwt}
    \end{subfigure}
    \begin{subfigure}{0.33\textwidth}
        \includegraphics[width=\linewidth]{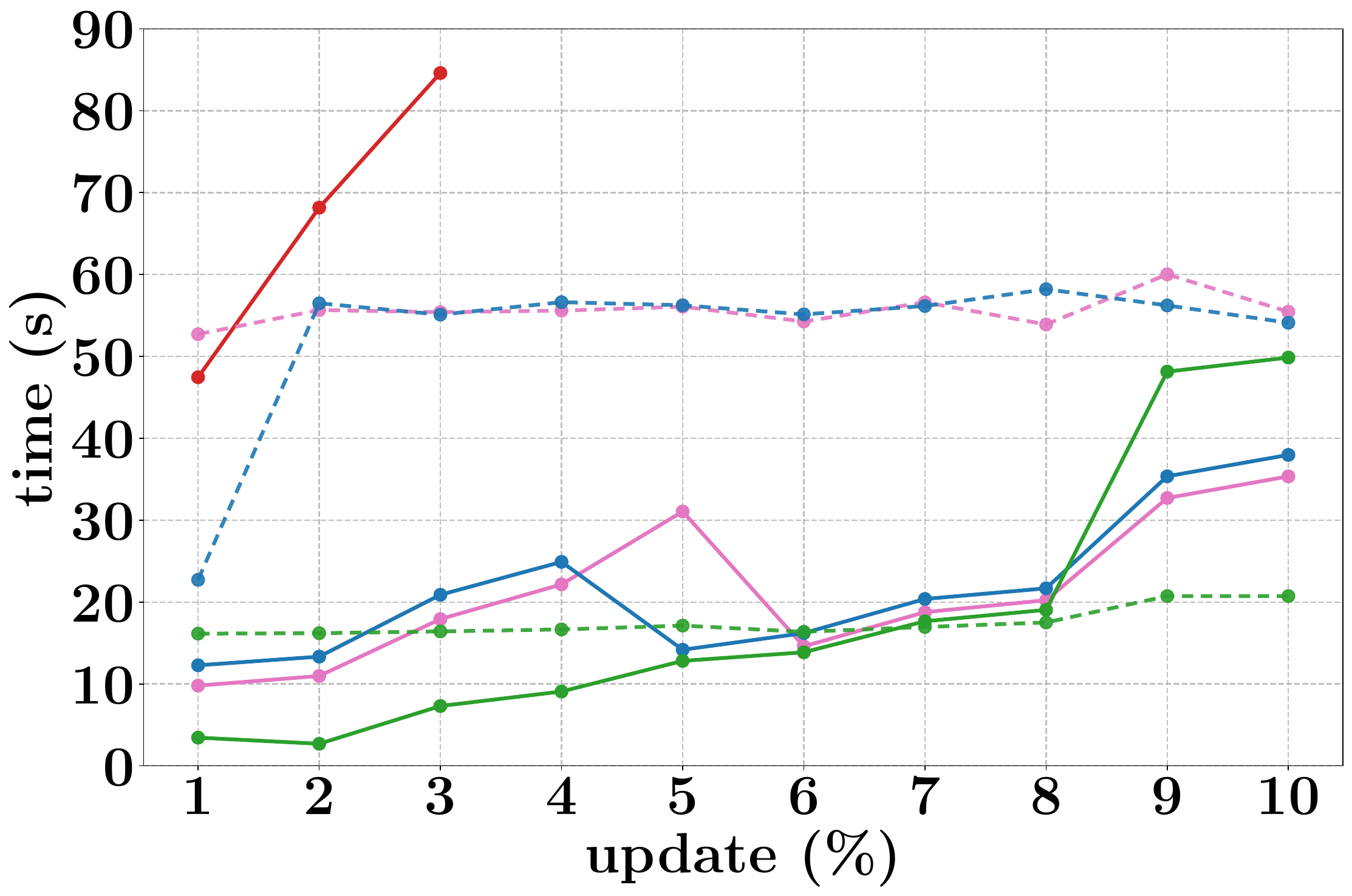}
        \caption{Wikiwt}
    \end{subfigure}

    \begin{subfigure}{0.33\textwidth}
        \includegraphics[width=\linewidth]{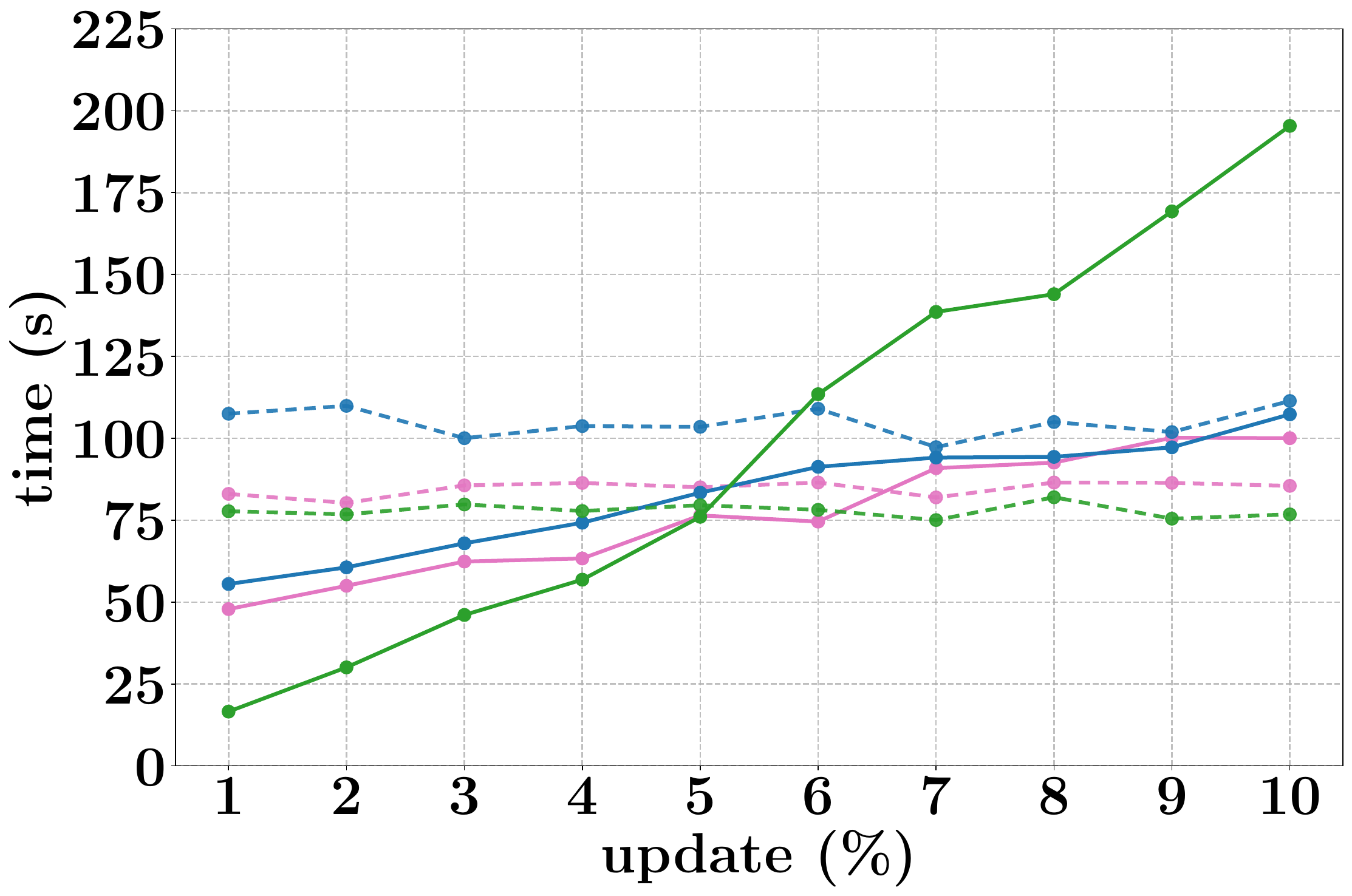}
        \caption{eu-2015-tpd }
    \end{subfigure}
    \begin{subfigure}{0.33\textwidth}
        \includegraphics[width=\linewidth]{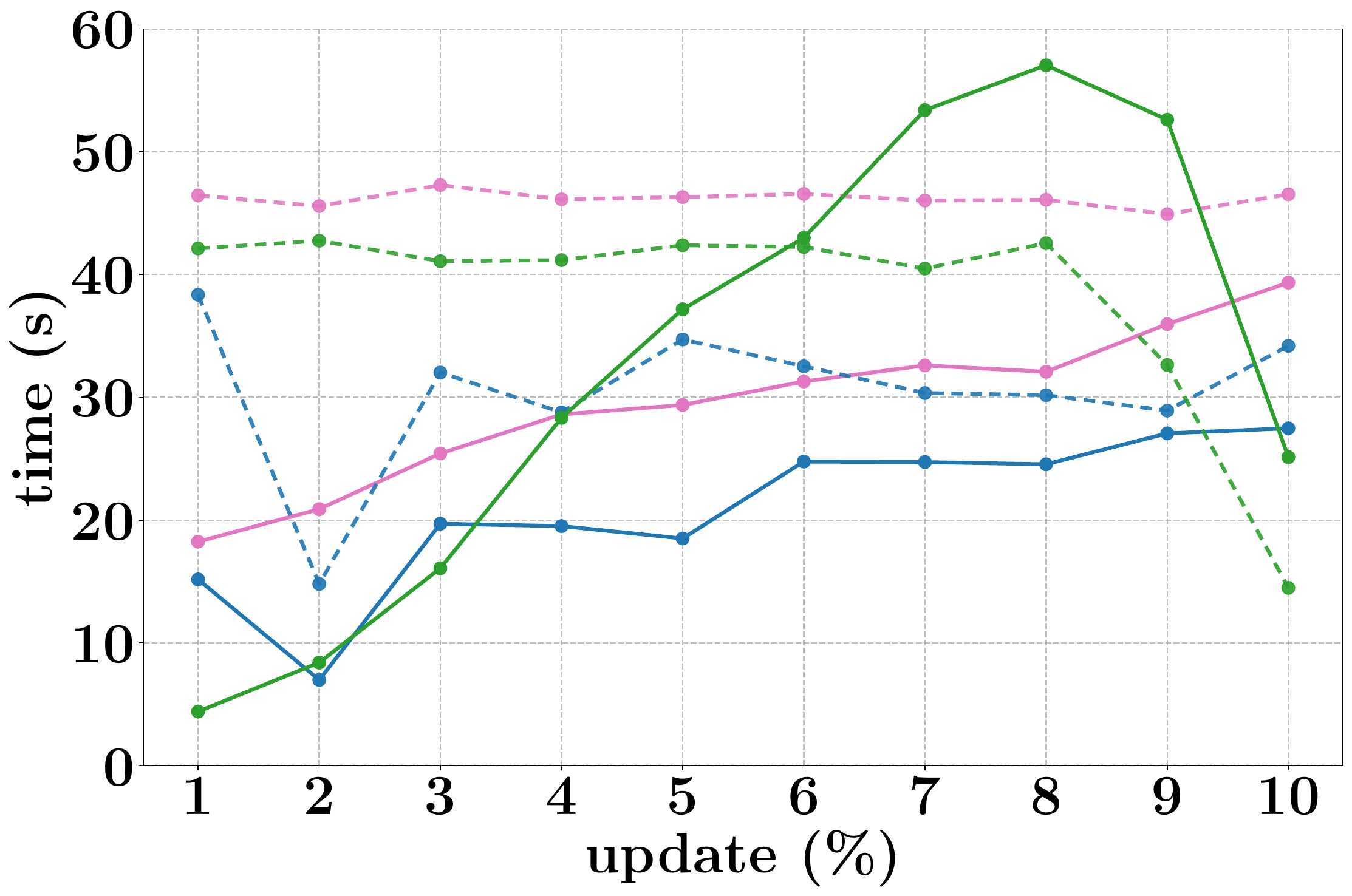}
        \caption{Orkutudwt}
    \end{subfigure}
    \begin{subfigure}{0.33\textwidth}
        \includegraphics[width=\linewidth]{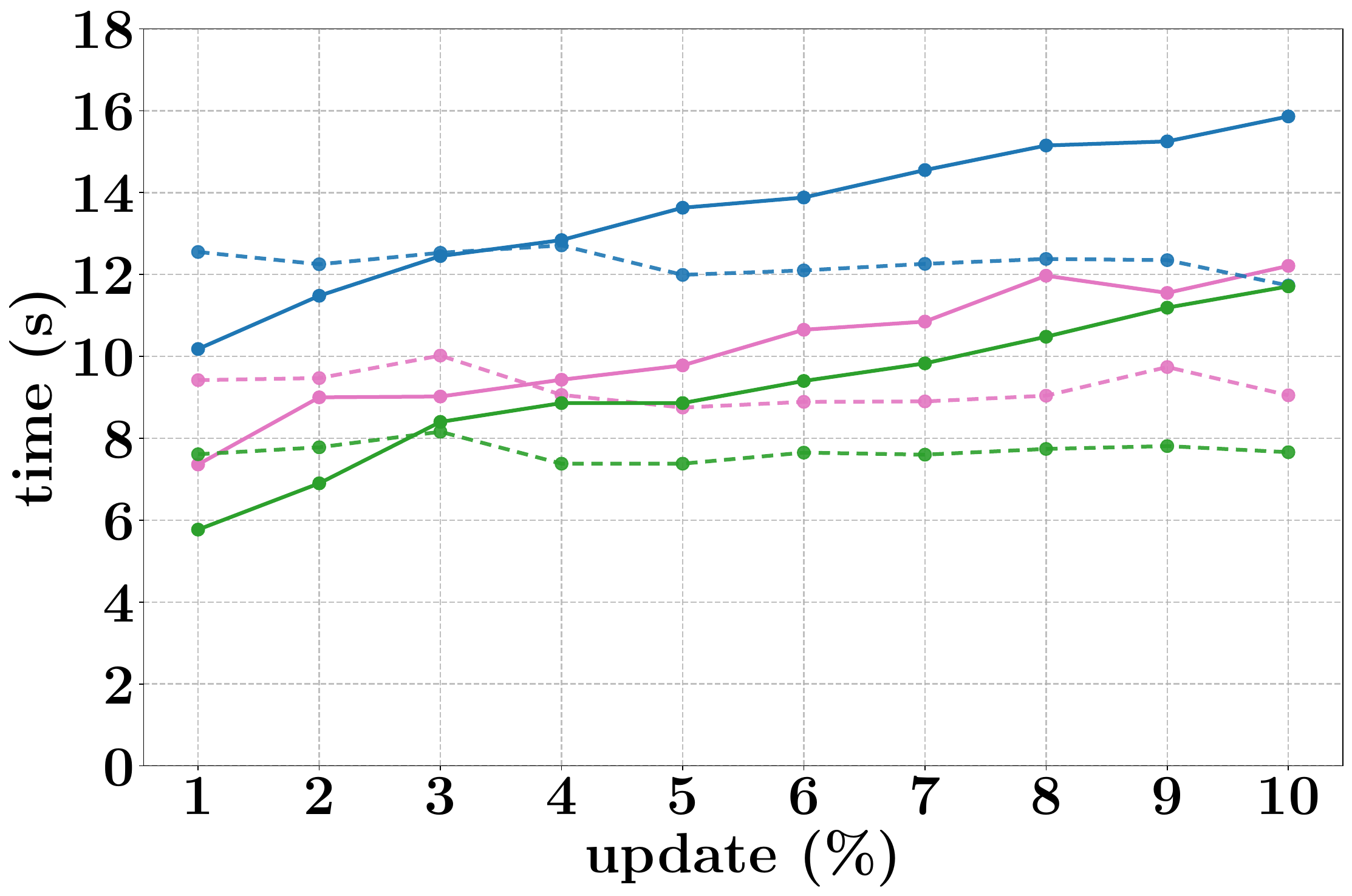}
        \caption{UK-2002}
    \end{subfigure}

\caption{
Performance on Mixed updates. Legend:
\protect\tikz[baseline=-0.6ex]\protect\node[fill=tabred,draw,circle,inner sep=2pt]{};
~alt-pp,\ 
\protect\tikz[baseline=-0.6ex]\protect\node[fill=tabgreen,draw,circle,inner sep=2pt]{};
~Push--Pull,\ 
\protect\tikz[baseline=-0.6ex]\protect\node[fill=tabblue,draw,circle,inner sep=2pt]{};
~Topology-driven,\ 
\protect\tikz[baseline=-0.6ex]\protect\node[fill=tabpink,draw,circle,inner sep=2pt]{};
~Data-driven.
\\Solid lines = dynamic, dashed lines = static.
}
\label{fig:set3}
\end{figure*}
In mixed updates, half of the edges in the batch have higher updated capacity, while the rest have lower updated capacity. The results are presented in Figure~\ref{fig:set3}. Such updates reflect the real-life scenarios where the updates are not purely incremental or decremental; they reflect the behaviour of the size of inputs rather than the type. 

For mixed updates, the \textsf{Push--Pull Streams (PP)} implementation maintains performance between its incremental and decremental trends, exhibiting strong gains at smaller update ratios across medium and large datasets. On \textsf{Stack Overflow}, PP achieves $7\times$ and $5.6\times$ improvements over \textsf{alt-pp} for $1$–$2\%$ updates, and $3.2\times$ and $2.2\times$ against static, while the topology-driven variant sustains on average $1.2\times$  up to $6\%$. Similarly, on \textsf{eu-2015-tpd} and \textsf{Orkutudwt}, PP records $4.7\times$ and $8.7\times$ improvements at $1\%$, respectively, while maintaining $2.6\times$ and $1.6\times$ average gains up to $4\%$ and $9\%$ against static recomputation.  

The \textsf{Data-Driven} variant offers steady advantages on smaller graphs. Against \textsf{alt-pp}, it delivers $2.4\times$ on \textsf{Flickr-Links}, $6.9\times$ on \textsf{LivJournalwt}, and $146\times$ on \textsf{Hollywood} on average till $10\%$ updates. On the other hand, \textsf{Topology-Driven} performs comparably with $3.5\times$ on \textsf{Pokecwt} and $154\times$ on \textsf{Hollywood} on average against \textsf{alt-pp}. Against static recomputation, the data-driven variant achieves $1.7\times$ on \textsf{Flickr-Links} and up to $1.5\times$ on \textsf{Hollywood} for $1\%$ updates, while PP attains $1.6\times$ on \textsf{LivJournalwt} till $4\%$ and $2.9\times$ on \textsf{Wikiwt} till $6\%$.  

Overall, the \textsf{data-driven} and \textsf{topology-driven} methods show similar behavior, with topology-driven slightly outperforming for higher update ratios, while PP dominates at smaller percentages on large graphs. 
\subsection{Observations}
Based on our experiments, we observe the following:
\begin{itemize}
    \item As the batch size increases, the time taken for dynamic processing generally increases for all three approaches.
\item Both the topology-driven and the data-driven approaches illustrate relatively similar behavior (same slope) across different batch sizes. 
\item For most datasets, the dynamic time is initially lower than the static time. This shows the benefit of running a dynamic processing. On the other hand, as the batch size increases, time for the dynamic processing also increases. Beyond a point, the static method outperforms the dynamic processing. This is expected because the overheads of the dynamic processing are overshadowed by the static processing when the number of updates is large. 
\item For larger graphs, the benefit of Push-Pull Relabel was visible, whereas it was inefficient for smaller graphs. Hence, assuming that the Min-cut remains mostly the same,  is more appropriate for larger graphs.
\item The topological approach dominates in certain cases, suggesting that the cost of worklist creation exceeds the overhead from idle threads, implying a high number of active vertices.
\end{itemize}

\section{Conclusion}\label{sec conclusion}
We presented efficient algorithms towards solving the dynamic \maxflow problem on GPUs. We developed the Dynamic Push Relabel algorithm, proved its correctness, and  implemented strategies: topology-driven and data-driven. We then extended the same methodology to the Dynamic Push Pull algorithm, implemented through the push–pull stream. From the experiments, we can conclude that Push-Pull implementation performs well for small-sized updates and large graphs. Data-Driven approach provides good performance for small and medium-sized graphs. Topology-Driven approach performs better for graphs with more active vertices.  Static algorithms start outperforming beyond a certain threshold. In the future, we would like to explore other graph algorithms and applications that may benefit from our approaches. We are also interested in exploring hardware specific optimizations. 




\bibliographystyle{ACM-Reference-Format}
\bibliography{references}

@String{Computing = "Computing" }

@ArtifactSoftware{R,
    title = {R: A Language and Environment for Statistical Computing},
    author = {{R Core Team}},
    organization = {R Foundation for Statistical Computing},
    address = {Vienna, Austria},
    year = {2019},
    url = {https://www.R-project.org/},
}

@article{ford1956,
  author    = {L. R. Ford Jr. and D. R. Fulkerson},
  title     = {Maximal flow through a network},
  journal   = {Canadian Journal of Mathematics},
  year      = {1956}
}

@article{edmonds1972,
  author    = {J. Edmonds and R. M. Karp},
  title     = {Theoretical improvements in algorithmic efficiency for network flow problems},
  journal   = {Journal of the ACM},
  year      = {1972}
}

@article{dinitz1970,
  author    = {E. A. Dinitz},
  title     = {Algorithm for solution of a problem of maximum flow in a network with power estimation},
  journal   = {Soviet Math. Dokl.},
  year      = {1970}
}

@article{karzanov1974,
  author    = {A. V. Karzanov},
  title     = {Determining the maximal flow in a network by the method of preflows},
  journal   = {Soviet Math. Dokl.},
  year      = {1974}
}

@article{goldberg1988,
  author    = {A. V. Goldberg and R. E. Tarjan},
  title     = {A new approach to the maximum flow problem},
  journal   = {Journal of the ACM},
  year      = {1988}
}

@article{anderson1995,
  author    = {R. Anderson and J. Setubal},
  title     = {A parallel implementation of the push-relabel algorithm for the maximum flow problem},
  journal   = {Journal of Parallel and Distributed Computing},
  year      = {1995}
}

@inproceedings{hong2008,
  author    = {Bo Hong},
  title     = {A Lock-Free Multi-Threaded Algorithm for the Maximum Flow Problem},
  booktitle = {Proceedings of the 23rd International Parallel and Distributed Processing Symposium (IPDPS)},
  year      = {2008},
  pages     = {1--8}
}

@INPROCEEDINGS{juntong,
  author={Luo, Juntong and Sallinen, Scott and Ripeanu, Matei},
  booktitle={2023 IEEE International Conference on Big Data (BigData)}, 
  title={Maximum Flow on Highly Dynamic Graphs}, 
  year={2023},
  volume={},
  number={},
  pages={522-529}, 
  doi={10.1109/BigData59044.2023.10386845}
}

@incollection{wu2012gems,
  author    = {Jing Wu and Zhiyuan He and Bo Hong},
  title     = {Efficient CUDA Algorithms for the Maximum Network Flow Problem},
  booktitle = {GPU Computing Gems Jade Edition},
  editor    = {Wen-mei W. Hwu},
  pages     = {55--66},
  publisher = {Elsevier},
  year      = {2012},
  isbn      = {9780123859631},
  url       = {https://www.sciencedirect.com/science/article/abs/pii/B9780123859631000058}
}

@article{Khatri2022,
  author    = {Jash Khatri and Arihant Samar and Bikash Behera and Rupesh Nasre},
  title     = {Scaling the Maximum Flow Computation on {GPUs}},
  journal   = {International Journal of Parallel Programming},
  year      = {2022},
  volume    = {50},
  number    = {5},
  pages     = {515--561},
  month     = dec,
  issn      = {1573-7640},
  doi       = {10.1007/s10766-022-00740-7},
  url       = {https://doi.org/10.1007/s10766-022-00740-7},

}

@INPROCEEDINGS{6569834,
  author={Nasre, Rupesh and Burtscher, Martin and Pingali, Keshav},
  booktitle={2013 IEEE 27th International Symposium on Parallel and Distributed Processing}, 
  title={Data-Driven Versus Topology-driven Irregular Computations on GPUs}, 
  year={2013},
  volume={},
  number={},
  pages={463-474},
  doi={10.1109/IPDPS.2013.28}}

@INPROCEEDINGS{he10,
  author={He, Zhengyu and Hong, Bo},
  booktitle={2010 IEEE International Symposium on Parallel \& Distributed Processing (IPDPS)}, 
  title={{Dynamically tuned push-relabel algorithm for the maximum flow problem on CPU-GPU-Hybrid platforms}}, 
  year={2010},
  volume={},
  number={},
  pages={1-10},
  doi={10.1109/IPDPS.2010.5470401}}

@inproceedings{brand23,
author = {van den Brand, Jan and Liu, Yang P. and Sidford, Aaron},
title = {{Dynamic Maxflow via Dynamic Interior Point Methods}},
year = {2023},
isbn = {9781450399135},
publisher = {Association for Computing Machinery},
address = {New York, NY, USA},
url = {https://doi.org/10.1145/3564246.3585135},
doi = {10.1145/3564246.3585135},
booktitle = {Proceedings of the 55th Annual ACM Symposium on Theory of Computing},
pages = {1215–1228},
numpages = {14},
location = {Orlando, FL, USA},
series = {STOC 2023}
}

\end{document}